\documentstyle[psfig,aps,prb,twocolumn]{revtex}

\begin{document}

\draft
\twocolumn[\hsize\textwidth\columnwidth\hsize\csname
@twocolumnfalse\endcsname

\title{Tunneling in the topological mechanism of superconductivity}

\author{A. G. Abanov}
\address{James Franck Institute of the University of Chicago,
5640 South Ellis Avenue, Chicago, Illinois 60637}

\author{P.B. Wiegmann}
\address{James Franck Institute
and Enrico Fermi Institute of the University of Chicago, 5640 South Ellis
Avenue, \\
Chicago, Illinois 60637 \\
and
Landau Institute for Theoretical Physics, Moscow, Russia}

\maketitle

\begin{abstract}
We compute  the two-particle matrix element and Josephson tunneling
amplitude in a two-dimensional model of topological superconductivity which
captures the physics of the doped Mott insulator. The
 hydrodynamics of topological electronic liquid consists of the compressible
charge sector and the incompressible chiral topological spin liquid. We  show
that  ground states differing by an odd number of particles are orthogonal and
insertion of two extra electrons is followed by the emission of soft modes of
the transversal spin current.
The orthogonality catastrophe makes the physics of
superconductivity drastically different from the BCS-theory but similar to
the physics of one-dimensional electronic liquids. The wave  function of a
pair  is dressed by soft modes. As  a result the  two particle matrix element
 forms a complex {\it d-wave representation} (i.e.\
changes sign under $90^o$ degree rotation), although the gap in the
electronic spectrum has
no nodes. In contrast to the BCS-theory the tunneling amplitude
has an asymmetric broad peak (much bigger than the gap) around the Fermi
surface. We  develop an operator algebra, that allows one to compute other
correlation functions.
\end{abstract}

]

\section{Introduction}
\label{Intro}

The phenomenon of ''superconductivity``---the existence of a metastable
quantum state with a current in a macroscopical system---manifests itself
as a particular
set of correlations in the ground state:

(i) Meissner effect
\begin{equation}
\label{1}
\lim_{|{\bf r}_1 - {\bf r}_2|\longrightarrow \infty}\langle{\bf j}_{\perp}
({\bf r}_1){\bf j}_{\perp}({\bf r}_2)\rangle=\frac{1}{(4\pi\lambda)^{2}},
\end{equation}
here ${\bf j}_{\perp}$
is the transversal current $\mbox{\boldmath $\nabla$}\cdot{\bf
j}_{\perp}=0$, and
$\lambda = \left(\frac{mc^2}{4\pi \rho_s e^2}\right)^{1/2}$ is the London
penetration depth and $\rho_s$ is the superfluid density;

(ii) Gap in the electronic spectrum:
singularity  of the one-particle
Green function
$\omega = \Omega({\bf p})$ closest to the origin in the $\omega$ plane
\begin{equation}
\label{2}
G(\omega, {\bf p})
= \langle c^{\dag}_{\sigma }(\omega, {\bf p})c_{\sigma }(\omega, {\bf
p})\rangle .
\end{equation}

(iii)
 Anomalous expectation value
\begin{equation}
\label{AnAv}
\langle N+2|c^{\dag}_{\uparrow}({\bf r}_1)
c^{\dag}_{\downarrow}({\bf r}_2)|N\rangle = \Delta({\bf r}_1-{\bf r}_2)\neq 0.
\end{equation}
The matrix element  between the ground states of
the system with $N$ and $N+2$ particles does not vanish in a macroscopic
system. It gives rise to Josephson tunneling, but not necessarily to an
off-diagonal long range order.

These three correlations describe very different sides of the phenomenon:
(i) hydrodynamics of an ideal liquid, (ii) a gap for one-particle
excitations, and (iii) a two-particle matrix
element.  Nevertheless, due to the mean field character of the BCS theory
all of them  turn out to be essentially the same---all these
quantities can be expressed in terms of one complex function $\Delta({\bf
r}_1-{\bf r}_2)$.
This misleading ''advantage`` of the BCS theory often allows one
to draw conclusions about the gap function (ii)
by looking  at the matrix element (iii) and vice versa.
However, the gap, Josephson current, and the penetration depth are essentially
different quantities: the first one characterizes the spectrum,  the second is
a  matrix element, determined also by the phase of the wave function, while
the third measures transversal current-current correlations.

In an electronic liquid where the interaction is strong, one also expects
to see a
difference between  dissimilar implementations of superconductivity
(\ref{1},\ref{2},\ref{AnAv}).
This difference becomes dramatic in the topological (anyon) mechanism of
superconductivity, where the entire effect of superconductivity is due to
peculiar quantum phases of  wave functions of the ground  state and low energy
excitations.

In this paper we study tunneling in two-dimensional topological
superconductors and show that:

(i)  like in the BCS-theory, although due to entirely different physics,
the Josephson
tunneling amplitude is proportional to the equal time matrix element
(\ref{AnAv}), and

(ii) the phase of the pair wave function  (\ref{AnAv})  depends on the
direction of the vector
${\bf r}_1-{\bf r}_2$ and forms  the complex  irreducible d-wave
representation of the group of rotations of the plane. The phase difference
between order parameters in points (1,2) and (3,4) is the twice  the angle
$\varphi $ between vectors ${\bf r}_1-{\bf r}_2$ and ${\bf r}_3-{\bf r}_4$
\begin{equation}
\label{3}
\frac{\Delta({\bf r}_1-{\bf r}_2)}{\Delta({\bf r}_3-{\bf r}_4)}
\sim e^{2i\varphi}.
\end{equation}
At the same time the gap function $\Omega({\bf
p})=\sqrt{\Delta_0^2+v^2p^2}$ has {\em no nodes}.

Moreover, the pair wave function can also be written as a product of the
BCS wave
function and  matrix element of the soft modes of the transversal spin current
in the coordinate space, averaged over the Fermi surface with the weight
$e^{i\arg({\bf k})}$
\begin{equation}
\label{w20}
\Delta({\bf r}) \sim  e^{-i2\arg ({\bf r})}\oint d{\bf k}_f
D_{{\bf k}_f}({\bf r})\Delta_{\rm BCS}({\bf k}_f,r),
\end{equation}
where
\begin{equation}
\label{BCS20}
\Delta_{\rm BCS}({\bf k}_f,{\bf r})\sim
\int kdk\, e^{i k \hat{{\bf k}}_f\cdot{\bf r}}
\frac{\Delta_0}{((\epsilon ({\bf k})-\mu)^2+\Delta_0^2)^{1/2}}
\end{equation}
is the BCS wave function of two particles with a relative momentum directed
along ${\hat {\bf k}}_f\equiv {\bf k}_f/k_f$ and the propagator of soft
modes is
given by
\begin{equation}
\label{w3}
D_{{\bf k}_f}({\bf r})=\frac{1}{k_fr}e^{i(\arg {\bf k}_f-\arg{\bf r})}\, .
\end{equation}

The factor 2 in the angular dependence
reflects the double degeneracy of {\it zero modes} and
eventually the spin of the electron. The  complexity of the tunneling amplitude
is a result of the violation of the time-reversal symmetry---an inherent
feature
of layered topological fluids with an odd number of layers. [In more
realistic, anisotropic three-dimensional systems or in the layered system
with even number of layers the time reversal symmetry is restored due to
alternating signs of parity breaking in consecutive layers
\cite{LZL,WTS1}.]

Below we consider the incommensurate case, where the dispersion
$\epsilon({\bf k})-\mu\sim v(|{\bf k}|-k_f)$ is generic and the Fermi
surface is a circle (Sec.\ \ref{TopSec}), and separately
the commensurate case on the square lattice near half filling
(Sec.\ \ref{OPCS}).

\begin{figure}
\centerline{\psfig{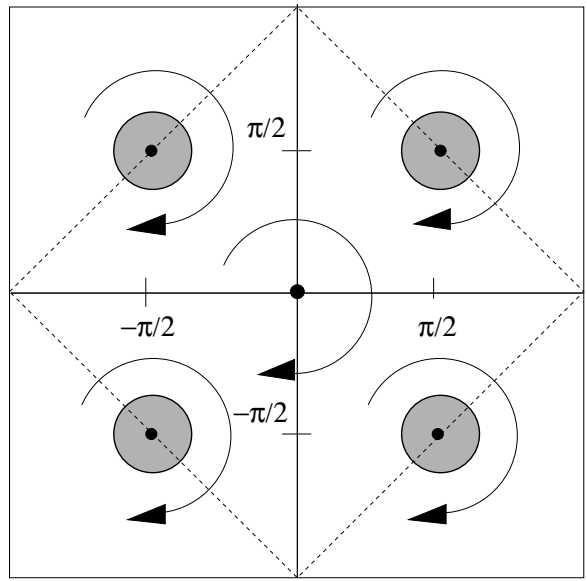}}
\vspace{0.5 cm}
\caption{The structure of the  pair wave function in momentum space:
$\Delta({\bf k})$  is maximal around four Fermi points ${\bf  k}_f
=(\pm\frac{\pi}{2},\pm\frac{\pi}{2})$ and consists of four
similarly oriented unit vortices around  each Fermi point. The phase of each
vortex is relative to the direction of the ${\bf k}_f$ as in the eq.
(\protect\ref{popo}).  It changes sign under $90^o$
rotation
$\Delta(-k_y,k_x)=-\Delta(k_x,k_y)$.}
\label{tunnampl}
\end{figure}

 Results for the commensurate case have been reported in
Ref. \cite{AW}. In this case the Fermi surface consists of four pockets around
${\bf k}_f=(\pm\pi/2,\pm\pi/2)$ with dispersion
$\epsilon({\bf k})-\mu\sim \sum_{{k}_f
=(\pm\frac{\pi}{2},\pm\frac{\pi}{2})}v(|{\bf k}|-k_f)$.
Formulae (\ref{w20},\ref{BCS20},\ref{w3}) in this case give
\begin{eqnarray}
\Delta({\bf R})
&\sim&
\frac{\sin\frac{\pi}{2}(X+Y)+i\sin\frac{\pi}{2}(X-Y)}
{X+iY}\Delta_{\rm BCS}({\bf R})\label{kl0}\\
&=&\frac{1}{X+iY}\sum_{{\bf k_f}
=(\pm\frac{\pi}{2},\pm\frac{\pi}{2})}
e^{-i\arg({\bf k_f}) + i\,{\bf k_f}{\bf R}}\Delta_{\rm BCS}({\bf R}),
\label{klr0}
\end{eqnarray}
where $(X,Y)$ are integer coordinates of the
lattice vector ${\bf R}\equiv{\bf r}-{\bf r}' $. The structure of the the
pair wave function
 in momentum space is shown in the
Fig.\ref{tunnampl}. Although the angular dependence  in (\ref{kl0}) is more
complicated than in the incommensurate case (\ref{w20}), it also realizes the
irreducible d-wave representation of the group of rotations of the square
lattice. In particular it changes sign under $90^o$ rotation
$\Delta(-Y,X)=-\Delta(X,Y)$.

The angular dependence of tunneling in anyon superconductors has been also
studied in Refs.\cite{R},\cite{L}.
One of the motivations of this work was the corner-SQUID-junction
experiment \cite{Jos} in which the relative phase of tunneling amplitudes
(\ref{AnAv}) on different faces of a single crystal of YBCO has been
measured. It is found to be $\pi$ in accord with d-wave superconductivity and
apparently in agreement with the topological  mechanism (\ref{3}).

These and many other anomalies in tunneling, transport, and the photoemission
spectrum are due to a single phenomenon: the orthogonality
catastrophe \cite{AndersonOrt}. In a strongly interacting environment the
addition of an extra particle to the system drastically changes its ground
state. As a result the overlap between the ground states  differing by an
{\it odd}
number of particles  vanishes in the macroscopical system. Contrary
ground states
which differ by an even number of particles (\ref{AnAv}) are not orthogonal
 (compare to Ref.\cite{AndersonChak}).

Tunneling involves a correlation between particles with different spins.
 Another goal of this paper is to develop a method to
treat the spin correlation.

This paper consists of four distinct parts. In Sec.\ \ref{TOC} we discuss
the  Josephson tunneling in the presence of orthogonality catastrophe. Then,
in Sec.\ \ref{TopSec}  we give a phenomenological description of the
topological mechanism of superconductivity.  We describe   hydrodynamics in
Sec.\ \ref{HydroSS} and include Fermi surface in the consideration in Sec.\
\ref{FermiSurface}. In Sec.\ \ref{EFT} we present a field theory model
(\ref{Pauli}) which yields  the phenomenological picture of Sec.\
\ref{TopSec}. We calculate tunneling amplitude and off-diagonal matrix
elements for the incommensurate case in Secs.\ \ref{es},\ref{ADT}
and for the nearly
half-filled (commensurate) case in Sec.\ \ref{copc}. Finally, in  
the Appendix
we describe a way to derive the model of topological superconductivity from a
t-J model, a canonical model of correlated electronic systems.  Although this
section is of interest in its own right, we present it separately as an
appendix for the sake of continuity of the text.

\section{Tunneling and orthogonality catastrophe}
\label{TOC}

Josephson tunneling is peculiar in the presence of an orthogonality
catastrophe. The Josephson current through the junction (at zero bias
voltage) between
two superconductors, of which one is conventional, is given by the well known
formula \cite{AmbBar}
\begin{eqnarray}
  I &=& - \,{\rm Im}\, i \sum_{kp}  T_{ kp} T_{-k-p} \frac{\Delta^*_1}{E_k}
 \nonumber \\
   &\times & \int_{0}^{\infty} \frac{d\omega}{2\pi}
   \left[ \frac{F(p,\omega)}{\omega-E_k-i\eta}
         -\frac{F(p,-\omega)}{\omega-E_k+i\eta} \right],
 \label{TUN}
\end{eqnarray}
where $T_{kp}$ is a transmission amplitude of the junction,
$E_k=\sqrt{|\Delta_1|^2+\epsilon_k^2}$ is the spectrum of the conventional
superconductor and $F(p,\omega)$ is the spectral function of the
superconductor of interest
\begin{eqnarray}
  F(p,\omega)  =  &2\pi & i\varepsilon^{\sigma\sigma^\prime}
  \sum \langle N|c_\sigma(p)|N+1\rangle
 \nonumber \\
    &\times &  \langle N+1|c_{\sigma^\prime}(-p)|N+2\rangle
  \delta(\omega-\epsilon_{p}).
 \label{SF}
\end{eqnarray}
Here the sum goes over all quantum  states with one extra particle
($\epsilon_{p}$ is the energy of an intermediate state). If the spectrum is
symmetric with respect to adding or removing a particle, i.e., $F(p,\omega)$
is an odd function of $\omega$,  we obtain
\begin{equation}
\label{TUNas}
  I = - 2\,{\rm Im}\, i \sum_{kp}  T_{kp} T_{-k-p} \frac{\Delta_1^*}{E_k}
  \int_{0}^{\infty} \, {\cal P}\frac{d\omega}{2\pi}
  \frac{F(p,\omega)}{\omega-E_k}.
\end{equation}
In the following we assume for simplicity that the transmission amplitude
$T_{kp}$ is strongly peaked at ${\bf k},\;{\bf p}$ close to the direction
normal to the junction. This simplification should not change the phase
dependence of the Josephson current although it can change the value of the
critical current. Assuming that the gap in the superconductor of interest
$\Delta_0$ is bigger than the one in the conventional superconductor
$\Delta_0\gg\Delta_1$ we obtain:
\begin{equation}
 \label{TUNassym}
   I \sim  |T|^{2} \nu_{0}|\Delta_1|\sum_{|{\bf p}_n|}
   \int_{0}^{\infty}\frac{d\omega}{\omega}
   \sin(\phi_0-\phi ({{\bf p}_n},\omega)) |F({\bf p}_n,\omega)|,
\end{equation}
where ${\bf p}_n$ is the component of  momentum normal to the surface of the
junction and averaging over ${\bf p}_n$ is determined by the actual form of
the transition amplitude. In this formula $\phi_0$ and $\nu_{0}$  are the
phase  and  density of states  of the conventional superconductor and $\phi
({\bf p},\omega)$ is the phase of the $F$-function (\ref{SF}) of the
superconductor of interest.

In the BCS theory the $F$-function has a peak at the gap
$\omega\sim\Delta_0$, such that the width  of the peak is also of the order
of the gap:
\begin{equation}
 \label{BCS1}
   \int dpF_{\rm BCS}(p,\omega) \sim
   \frac{\Delta_0}{\sqrt{\omega^{2}-\Delta^2_0}}.
\end{equation}
The peak gives the major contribution to the integral (\ref{TUNassym}). It
selects a characteristic energy of the intermediate state $\epsilon_{q}\sim
\Delta_0$ and gives rise to the traditional BCS picture of tunneling: a
pair decays into two electrons while tunneling, so electrons tunnel
independently. Short time processes (at $\omega\sim\epsilon_f$) do not
contribute to the integral (\ref{TUNassym}).

The situation is drastically different in the orthogonality catastrophe
environment \cite{AndersonChak}. We will show that in a topological
superconductor an individual matrix element $\langle
N|c_{\sigma}(p)|N+1\rangle$
acquires an additional factor $1/Lk_f$, where $L$ is the size of the system
and therefore vanishes in a macroscopical sample, i.e., the ground states with
$N$ and $N+1$  particles are almost orthogonal. Nevertheless, the tunneling,
i.e., a matrix element between states with $N$ and $N+2$ particles is nonzero
due to a large number of low-energy intermediate states contributing to the
sum (\ref{SF}). A result of this is that the spectral function (\ref{SF})
acquires an additional factor $\omega/\epsilon_f$. In contrast to BCS at
$\omega\gg\Delta_0$ we have
\begin{equation}
\label{Eq}
\int dpF(p,\omega) \sim \frac{\Delta_0}{\omega}
\left(\frac{\omega}{\epsilon_f}\right) \sim \frac{\Delta_0}{\epsilon_f}.
\end{equation}
Therefore, the characteristic scale of spectral function is shifted to
the ultraviolet and becomes of the order of the Fermi energy---much larger
than the scale of the gap [A similar phenomenon occurs in momentum
space. In the Sec.\ \ref{ADT} we show that   the pair wave function $\Delta
(k)$
has a long tail well away from the Fermi surface (see also Fig.\
\ref{ordpar})]:
the integral (\ref{TUNassym})
is saturated by $\omega \sim \epsilon_f\gg\Delta_0$. This means that a pair
remains
intact during the tunneling and the tunneling amplitude is determined by
the {\it equal time} value of $F(p,t=0)$, i.e., by the matrix element
$\Delta({\bf r})$ of
an instantaneous creation of a pair (\ref{AnAv}).
Let us  notice that the correction to the spectral function in the
eq. (\ref{Eq})---$(\omega/\epsilon_f)^\alpha$ at $\alpha=1$ is just marginal.
Were
$\alpha$ be less than 1 the time of the tunneling would be of the order of
$\Delta_0^{-1}$
(A similar phenomenon has been discussed by Chakravarty and Anderson in the
context of interlayer tunneling in cuprate superconductors
\cite{AndersonChak}).

In the coordinate representation the Josephson current is
\begin{eqnarray}
\label{TUN2}
I \sim \,{\rm Im}\, e^{-i\phi_o} \Delta({k_f{\bf n}})\sim\sin(\phi_0-\phi
(k_f{\bf n},\omega=0)).
\end{eqnarray}

In the corner-SQUID-junction geometry \cite{GeshLar,Jos} one
can directly measure the difference of phases of $\Delta$ between two faces of
the superconducting crystal.

The instantaneous character of tunneling and the power laws (\ref{Eq}) are
known  from one-dimensional electronic systems where the orthogonality
catastrophe comes to its own. In addition, in 2D it also leads to the
angular dependence of the two-particle amplitude (\ref{AnAv},\ref{w20}).

The direct tunneling current is also strongly affected by
the orthogonality catastrophe:
\begin{equation}
\label{IV1}
I_{\rm dir}=2|T|^2\nu_{0}\int_{0}^{eV}\frac{d\omega}{2\pi}
\int \frac{d {\bf k}}{(2\pi)^2}\,{\rm Im}\, G(\omega,{\bf k}).
\end{equation}

Elsewhere  we will show that due to the orthogonality catastrophe the 
one-particle Green function acquires a branch cut, rather than a pole on the
threshold. As a result the density of states in a topological superconductor
just above the gap is suppressed by the factor $\sqrt {\omega-\Delta_0}$.
This leads to  a suppression of direct current close to the threshold bias
voltage. With a logarithmic accuracy we will obtain almost linear I-V
behavior
\begin{equation}
 \label{IV2}
   I_{\rm dir}\sim (V-\Delta_0).
\end{equation}
This is to be compared to  the  BCS-theory result:
$I_{\rm dir}\sim (V-\Delta_0)^{1/2}$.

\section {Topological Mechanism of Superconductivity}
\label{TopSec}

\subsection{Hydrodynamics}
\label{HydroSS}

The mechanism of  topological superconductivity is a
generalization to higher dimensions of the Peierls-Fr\"{o}hlich phenomenon,
known from one-dimensional electron-phonon systems
\cite{Froehlich54,Peierls,L1,MFappr,WTS1,WTS2}. It  inherits major features
from electronic physics in one dimension. We attempt to present it here
without addressing a particular model but accepting some minimal assumptions.
Later in the Appendix we develop a microscopical model to justify and
illustrate the phenomenological picture.

{\it (i) Zero Modes and Topological Instability.} Let us consider an
electronic liquid where
the interaction between electrons is mediated by an electrically neutral
bosonic field, that can form a point-like spatial topological configuration
(soliton). Let us suppose, that in a sector with zero topological charge the
electronic spectrum has a gap $m$. Assume now that in the presence of  a static
soliton the electronic spectrum differs from an unperturbed one by an
additional state just at the top of the valence band or within the gap---a
so-called zero mode or a midgap state \cite{ZerModeRef}. If the zero mode is
separated from the spectrum, its wave function is localized around the core
of the topological defect. In case when the level is attached to a band, the
wave function decays as a power law away from the center of the soliton. A
general argument \cite{GlobAnom} suggests that the midgap state always has an
{\it even} degeneracy. This degeneracy leads eventually to a proper flux
quantization and below it is assumed to be twofold.

Now let us add an even number of extra electrons with a concentration
$\delta$ into the system. They may occupy a new state at the Fermi level of
the conduction band. It costs the energy of the gap plus the Fermi energy
$m+\mu_f$ per particle, where $\mu_f$ is a chemical potential. Alternatively,
the system may  create a topological configuration and a number of zero modes
in order to accommodate all extra particles. The energy of this state is the
soliton mass  plus exponentially small corrections due to the interactions
between zero modes. If the latter energy is less than
$\mu_f$ then every two extra
electrons  added to the system  create a soliton  and then completely fill  a
zero mode, rather than occupy  the Fermi level of the state with zero
topological charge. As a result the total number of solitons in the ground
state is equal to half of the total number of electrons in the system.

Formally it means that, contrary to the Landau Fermi-liquid picture, the
expansion of the energy in small smooth variation of chemical potential
$\delta\mu ({\bf r})$  has a non vanishing linear term in $\delta\mu$:
\begin{eqnarray}
   \delta E(\mu)
   = &\int & \delta\mu({\bf r}){\bar \rho}\,d{\bf r}
   + \int\delta\mu({\bf r})K(\mu,{\bf r}-{\bf r}')
   \frac{F({\bf r}')}{2\pi}\,d{\bf r}d{\bf r}'
 \nonumber \\
   &+& {\cal O}(\delta\mu^2).
 \label{linearterm}
\end{eqnarray}
Here ${\bar \rho}$ is the electronic density,
$F({\bf r})/2\pi$ is a density of
a topological charge and $K(\mu,{\bf r})$ is some kernel.
The linear term in chemical potential is  known as
Chern-Simons term.

The minimum of energy is achieved if the variation of density is followed by
the variation of the topological charge
\begin{equation}
 \label{flux1}
   \rho({\bf r})-\bar{\rho}=\int K(\mu,{\bf r}-{\bf r}')
   \frac{F({\bf r}')}{2\pi}d{\bf r'}.
\end{equation}
While doping, electrons  create and occupy zero mode states to
minimize their energy, thus giving a non zero value to the topological charge.
Due to the twofold degeneracy of zero mode states
\begin{equation}
   \int K(\mu,{\bf r})\,d{\bf r}=2
\end{equation}
so a number of extra particles $\delta$ gives rise to a flux
\begin{equation}
 \label{constraint}
   \delta=2\int\frac{F({\bf r})}{2\pi}d{\bf r}.
\end{equation}

We refer to this phenomenon as {\it topological instability}. Once zero mode
states are occupied, residual interaction between them lifts the degeneracy, so
that zero mode states form a narrow band. This band is always
completely filled and is detached from the rest of the spectrum by some gap
$\Delta_0$.

This is already sufficient to conclude about superconductivity --- the
chemical potential always lies in a gap. The following
arguments are borrowed from  Fr\"{o}hlich's paper \cite{Froehlich54}. The
position of the topological excitation is not fixed relative to the
crystal lattice. Therefore, a pair of electrons bound to a topological
excitation can easily slide through the system (and therefore carry electric
current).  It slides unattenuatedly, since the state is completely
filled and  is separated by the gap from the unoccupied electronic states. As
a result, the low energy physics of density fluctuations is described by the
hydrodynamics of a liquid of zero modes:
\begin{equation}
 \label{hydrodynamics}
   H=\frac{1}{2\bar{\rho}}[v_{0}^{2}(\rho-\bar{\rho})^2+{\bf j}^2]
\end{equation}
where
$\bar{\rho},\,v_{0}$
are the average density and the velocity of the sound mode, ${\bf
j}=\rho{\bf v}$ is the current and ${\bf v}$ is the velocity
\begin{equation}
 \label{velocity}
   [\rho({\bf r}),{\bf v}({\bf r}')]=-i\mbox{\boldmath $\nabla$}_r
   \delta({\bf r}-{\bf r}').
\end{equation}
In dimensions higher than one, where the pinning effects are not that
important, the eqs. (\ref{hydrodynamics},\ref{velocity}) already imply
the Meissner effect (\ref{1}) and superconductivity \cite{WTS1,WTS2}.

It is convenient to re-parameterize densities and currents in
terms of charge displacement ${\bf u}$:
\begin{equation}
 \label{u}
   \rho-\bar{\rho}=-\mbox{\boldmath $\nabla$}{\bf u},\;\;\;
   {\bf j}=\partial_t{\bf u},
\end{equation}
so that
\begin{equation}
 \label{current}
   [j_i({\bf r}),u_k({\bf r}')]
   =-i\bar{\rho}\delta_{ik}\delta ({\bf r}-{\bf r}').
\end{equation}

Let us stress that this mechanism is very different from the mechanism where an
electronic pair is localized by a polaron. In contrast, in the topological
mechanism the electric charge of a pair is only partially localized at the core
of the topological soliton. Although the number of zero modes is equal to the
topological charge of the soliton,  a part of the charge is smoothly
distributed throughout the rest of the system.

{\it (ii)  Single-valuedness and gauge invariance.}
The electronic wave function is single-valued.  This important property may be
reformulated in terms of gauge invariance. In two spatial dimensions a
topological density can be described by means of a gauge field ${\bf A}$:
\begin{equation}
 \label{A}
   F({\bf r})=\mbox{\boldmath $\nabla$}\times {\bf A}.
\end{equation}
The single-valuedness of the electronic wave function means that electrons are
neutral with respect to this field: the electronic operator does not transform
under the gauge transformation ${\bf A}\rightarrow{\bf A}+\mbox{\boldmath
$\nabla$}\Lambda $, i.e.,
\begin{equation}
 \label{rho+}
   [ A_i({\bf r}), u_k({\bf r}')]=0;  \;\;\;
   [ A_i({\bf r}), j_k({\bf r}')]=0.
\end{equation}

Let us note that, despite of what eq.\ (\ref{flux1}) suggests, the vector
potential $A_i$ is not equal to  the displacement $\varepsilon^{ik}u_k$ but
differs from it by a singular gauge transformation.

In fact the wave functions of zero modes represent conformal blocks of
some conformal field theory.

{\it (iii) Incompressible chiral spin liquid.}  The hydrodynamics of charge
degrees of freedom alone is not sufficient to draw conclusions about fermionic
matrix elements. They are determined also by the distribution of electronic
spin within the zero mode. Unless there are solitons
 which create zero modes in
spin sector, the compressibility with respect to density modulations
generally results in incompressibility with respect to spin modulations.
Moreover, the spin liquid in the singlet sector is a topological liquid.
In the next section we show that on a very general basis, spin displacements
${\bf u}_s$ defined as $S^3=-1/2\mbox{\boldmath $\nabla$} {\bf u}_s$
obey anomalous commutation relations
\begin{equation}
 \label{spind}
   \left[{\bf u}_s ({\bf r})\times {\bf u}_s({\bf r}')\right]
   =2i\pi\delta ({\bf r}-{\bf r}')
\end{equation}
in addition to standard  $[j_{i,s}({\bf r}),u_{k,s}({\bf r}')]
=-i\bar{\rho}\delta_{ik}\delta ({\bf r}-{\bf r}')$.
This property determines rotational properties of a local
singlet---it has $l=2$ angular momentum---and  eventually determines the
phase of the tunneling amplitude (\ref{3}).

By combining charge and spin parts of the hydrodynamics
(\ref{hydrodynamics}, \ref{spind})  we
obtain the hydrodynamics of topological superconductivity:
\begin{eqnarray}
   {\cal L}&=&{\cal L}_c+{\cal L}_s,
 \label{CS1} \\
   {\cal L}_c&=&\frac{1}{2\bar{\rho}}((\partial_t{\bf u})^2-v_{0}^{2}
   (\mbox{\boldmath $\nabla$}{\bf u})^2),
 \label{Lcharge} \\
   {\cal L}_s &=&2\pi{\bf u}_s\times\partial_t {\bf u}_s
   +\frac{1}{2\bar{\rho}}((\partial_t{\bf u}_s)^2
   -v_{0}^2(\mbox{\boldmath $\nabla$}{\bf u}_s)^2).
 \label{Lspin}
\end{eqnarray}
The hydrodynamics consists of two independent fluids: a
compressible charged liquid
\begin{eqnarray}
 \label{liquer}
   \langle u^\parallel(\omega,{\bf k}),u^\parallel(-\omega,-{\bf k})\rangle
   &=& \frac{\bar{\rho}}{\omega^2-(v_0 {\bf k})^2},
 \nonumber\\
   \langle u^\perp(\omega,{\bf k}),u^\perp(-\omega,-{\bf k})\rangle
   &=& \frac{\bar{\rho}}{\omega^2},
\end{eqnarray}
and an incompressible topological (chiral) spin liquid
\begin{eqnarray}
   \langle u^\parallel_s(\omega,{\bf k}),
   u^\parallel_s(-\omega,-{\bf k})\rangle
   &=& -\frac{1}{\vartheta^2\bar{\rho}},
 \nonumber\\
  \langle u^\perp_s(\omega,{\bf k}),u^\perp_s(-\omega,-{\bf k})\rangle
   &=& -\frac{1}{\vartheta^2\bar{\rho}}
   (1-\frac{v_0^2{\bf k}^2}{\omega^2} ),
 \label{liquer2} \\
   \langle u^\parallel_s(\omega,{\bf k}),u^\perp_s(-\omega,-{\bf k})\rangle
   &=& \frac{i}{\vartheta\omega},
 \nonumber
\end{eqnarray}
where $\vartheta=4\pi$ and $u^\parallel$, $u^\perp$ are the longitudinal and
transversal parts of the displacement $ u_i({\bf
k})=\frac{k_i}{k}u^\parallel+\frac{\varepsilon^{ij}k_j}{k}u^\perp$.
Eqs.\ (\ref{liquer}) imply compressibility of the charge fluid and the
Meissner effect. The absence of the sound mode in the spin liquid
(\ref{liquer2}) is the consequence of the  topological mass generated by the
Chern-Simons term in (\ref{Lspin}).

{\it (iv) Edge Spin Current.}
One of the direct consequences of the chiral nature of the spin
liquid  is that the spin liquid generates spin edge current. Indeed,
the spin sector of the hydrodynamics (\ref{Lspin}) is equivalent to the
hydrodynamics of the FQHE fluid (see e.g.,
\cite{Laughlin88FQHEhTc,FQHEelectrodynamics}). Similar to the FQHE the spin
excitations are suppressed in the bulk but develop a spontaneous spin edge
current with the level $k=2$ current algebra. Let us stress that the spin
edge current is the only hydrodynamical manifestation of spontaneous parity
breaking. Contrary to a number of claims scattered through the literature,
the spontaneous parity breaking is invisible in the charge sector even for
the systems with an odd number of layers.

\subsection{The field theory}
\label{EFT}

Let us realize the commutation relation
(\ref{velocity},\ref{current},\ref{rho+},\ref{spind}) by means of an electron
operator:
\begin{eqnarray}
   \rho &=& c^\dagger c;\;\;\;\;\;\;
   {\bf j}=-ic^\dagger \mbox{\boldmath $\nabla$} c;
 \nonumber \\
   S^3 &=& \frac{1}{2} c^\dagger\sigma^3 c; \;\;\;\;
   {\bf j}_s=-\frac{1}{2}ic^\dagger\sigma^3\mbox{\boldmath $\nabla$} c.
\label{c}
\end{eqnarray}
The topological constraint (\ref{constraint}) reduces the Hilbert space of 
the low-energy sector of the theory. Instead of using a projected electronic
operator $c_\sigma$ we introduce a ''spinned`` fermion operator
$\psi_\sigma$ in which terms the field theory which takes topological
constraint (\ref{constraint}) into account is local.
%\begin{equation}
% \label{unw}
%   \psi_\sigma ={\cal V}_\sigma  c_\sigma .
%\end{equation}
%The vertex operator
%${\cal V}_{\sigma}({\bf r})$
%creates a flux quantum in the ground state with one extra particle with
%a spin $\sigma$\cite{Mand}:
%\begin{eqnarray}
%   [F({\bf r}'),{\cal V}_\sigma({\bf r})]
%   & = & 2\pi  {\cal V}_\sigma ({\bf r})\delta({\bf r}-{\bf r}').
% \label{AbVert}
%\end{eqnarray}
We discuss the relation between the physical electron operator
$c_\sigma$ and the unphysical gauge non-invariant operator $\psi_\sigma$
in the Secs.\ \ref{FermiSurface},\ref{VO}.
In $\psi$ representation the density of electrons is
%and the current are:
\begin{equation}
 \label{psi}
   \rho = \sum_\sigma c^\dagger_\sigma c_\sigma
        =\sum_\sigma\psi^\dagger_\sigma \psi_\sigma .
%    \;\;\;\;
%   {\bf j} = \psi^\dagger_\sigma(-i\mbox{\boldmath $\nabla$}+{\bf A})
%   \psi_\sigma.
\end{equation}
A standard example of the theory which exhibits the
topological instability and
therefore superconductivity is the Dirac Hamiltonian
\begin{equation}
 \label{Pauli}
   H = \psi^\dagger_{\sigma}\mbox{\boldmath $\alpha$}
       \big(-i\mbox{\boldmath $\nabla$}
   +{\bf  A}\big)\psi_{\sigma}
  +\Delta_0
\psi^\dagger_{\sigma}\beta\psi_{\sigma}-\mu\psi^\dagger_{\sigma}\psi_{\sigma}.
\end{equation}
where $\mbox{\boldmath $\alpha$}=(\alpha_x,\alpha_y)$ and $\beta$ are
$2\times 2$ Dirac
matrices:
$\{\alpha_x,\alpha_y\}=0,\;\beta=-i\alpha_x\alpha_y$. This Hamiltonian has
$\frac{2}{2\pi}\int F({\bf r})d{\bf r}$ zero
modes (the flux is directed up).  Wave
functions of zero modes are:
\begin{eqnarray}
 \label{zm0}
   \Phi({\bf r}) &=& e^{-i\int^{\bf r} {\bf A}({\bf r}')\cdot d{\bf r}'
    -\beta \int^{\bf r}{\bf A}({\bf r}')\times d{\bf r}'}\Phi_0({\bar z}),
 \\
    \beta \Phi_0({\bar z}) &=& -\Phi_0({\bar z})
 \nonumber
\end{eqnarray}
where $\Phi_0$ is any polynomial of degree
$\frac{1}{2\pi}\int F({\bf r})d{\bf r}-1$ \cite{AharonovCasher}.
The fact of existence of zero modes implies that the energy at
$\mu=0$ has a linear
term (\ref{linearterm}) in chemical potential
$K(\mu=0,{\bf r})=2\delta({\bf r})$.
Other models of topological superconductivity were discussed in
Refs.\cite{WTS1,WTS2}.

Let us comment on the relation between topological mechanism of
superconductivity and superconductivity in the system of anyons
\cite{L1,MFappr}.
The models become very close after projection onto the low
energy sector, where the relation (\ref{flux1}) is treated as a constraint
rather than
as a result of minimization of the energy. The projection can be done by
introducing a Lagrangian multiplier $A_0$ for the relation (\ref{flux1})
and commutation relations
$\left[A_x({\bf r}),A_y({\bf r}')\right]=2\pi i K^{-1}({\bf r}-{\bf r}')$,
i.e.\
by  adding the Chern-Simons term  with the kernel $K$ to the
Lagrangian.   In anyon model the kernel
$K({\bf r})$ is replaced by $2\delta ({\bf r})$ (even at $\mu\neq 0$),
so that the relation
between topological charge
and the density becomes local
$\rho({\bf r})=2\frac{F({\bf r})}{2\pi}$.
This simplification results in a
generation of transversal electric currents or ''internal`` magnetic field
by light or by inhomogeneous electric charge. Due to the same reason,
the Meissner effect in the anyon model
{\it per se} exists only at zero frequency, zero momentum, zero
temperature, infinitesimal magnetic field etc. These unphysical
consequences originate from
the topological constraint (\ref{constraint}). To avoid them one must determine
the kernel $K(\mu,{\bf r})$ self-consistently.

The topological mechanism of superconductivity takes place in certain
models of 2D doped Mott insulators.  We point out
several steps involved in the derivation of the Dirac Hamiltonian
(\ref{Pauli}) in Appendix, while here we would
like to give a physical interpretation
of the vector potential ${\bf A}$ and its flux ${\bf F}$.
The vector potential $A_i$ is
the phase of the hopping amplitude  of an electron in direction $i=x,y$.
Its flux is the phase of the total hopping amplitude along a closed path.
On a square lattice at small doping it is
\begin{equation}
 \label{chirality1}
   F({\bf r}) =-\,{\rm Im}\, \ln \,\mbox{tr}\,{\bf W},
\end{equation}
where the chirality ${\bf W}$ is defined  as
\begin{eqnarray}
   {\bf W} &=& (\frac{1}{2}+\mbox{\boldmath $\sigma$}\cdot {\bf S}({\bf r}))
   (\frac{1}{2}+\mbox{\boldmath $\sigma$}\cdot {\bf S}({\bf r} +{\bf e}_x))
 \nonumber \\
   &\times & (\frac{1}{2}+\mbox{\boldmath $\sigma$}\cdot {\bf S}({\bf r}
   +{\bf e}_y +{\bf e}_x))
 \label{chirality2}
\end{eqnarray}
and ${\bf r},{\bf r}+{\bf e}_x,{\bf r}+{\bf e}_x+{\bf e}_y$ are three
consecutive points of a lattice and $\mbox{\boldmath $\sigma$}$ are auxiliary
Pauli matrices.

\subsection{Hydrodynamics from the mean-field approximation}
\label{H}

The hydrodynamics of a superfluid (\ref{Lcharge}) can easily be
obtained from the
model (\ref{Pauli}).
Let us see how the energy (\ref{linearterm}) of a spin singlet state
changes under smooth
variations of ``electric'' ${\bf E}=\partial_t{\bf A}-\mbox{\boldmath
$\nabla$}\delta\mu$ and
``magnetic'' $F=\mbox{\boldmath$\nabla$}\times{\bf A}$ fields. To keep
track of spin variations
we  add an external field ${\bf {\cal A}}^3$ to the Hamiltonian (\ref{Pauli})
$i\mbox{\boldmath $\nabla$}-{\bf A}\rightarrow i\mbox{\boldmath
$\nabla$}-{\bf A}-{\bf {\cal
A}}^3\sigma_3$. In Gaussian approximation the result  consists of two
separate parts---spin
and charge. In the Coulomb gauge $\mbox{\boldmath $\nabla$}{\bf A}=0$ the
density of energy is
\begin{eqnarray}
 \label{mf}
   \delta E&=& \delta E_c+ \delta E_s,
 \label{Lcs}\\
    \delta E_c&=&\frac{\Pi_0}{2}{\bf E}^2 +
    \frac{\Pi_\perp}{2}F^2
   + \delta\mu\big( \frac{K}{2\pi} \mbox{\boldmath $\nabla$}\times{\bf a}
   -\delta{\rho}\big) ,
 \label{Lcharge1}\\
   \delta E_s &=& \frac{\Pi_0}{2}(\partial_t {\bf {\cal A}}^3)^2
   + \frac{\Pi_\perp}{2}(\mbox{\boldmath $\nabla$}\times{\bf {\cal A}}^3)^2
   + \frac{K}{2\pi}{\bf {\cal A}}^3\times\partial_t{\bf {\cal A}}^3 ,
 \label{Lspin1}
\end{eqnarray}
where polarization operators
\begin{eqnarray}
   \Pi_0 &=& \omega^{-2}\langle j_{\parallel}(\omega,{\bf k})
   j_{\parallel}(-\omega,-{\bf k})\rangle  ,
 \nonumber \\
   \Pi_\perp &= & k^{-2}
   \langle j_\perp (\omega,{\bf k}) j_\perp (-\omega,-{\bf k})
   -j_{\parallel}(\omega,{\bf k})j_{\parallel}(-\omega,-{\bf
   k})\rangle ,
 \nonumber \\
   K &= & i\omega^{-1}\langle j_\parallel(\omega,{\bf k})
   j_\perp (-\omega,-{\bf k}\rangle
  \label{pi}
\end{eqnarray}
are transversal current-current correlators of free Dirac massive fermions
at ${\bf
k},\omega\rightarrow 0$. Since the particles are massive  the propagators
(\ref{pi}) are  known
to be
$\Pi_0= \Pi_\perp v_0^{-2}\rightarrow \mbox{const}$ and
$K\rightarrow 2$ .

To obtain the hydrodynamics (\ref{CS1}) one must rewrite this result in
terms of
spin and charge displacements. Minimization over $\delta\mu$ gives the
relation
(\ref{flux1}) for the charge sector. Substituting this relation into
(\ref{Lcharge1})
gives the hydrodynamics of the ideal liquid (\ref{Lcharge}). In
the spin sector (\ref{Lspin1}) the story is different. The total spin is
kept to be zero.
Therefore, the Chern-Simons term remains in the spin sector and gives the
hydrodynamics of an incompressible spin liquid. Writing
$\partial_t{\bf u}_s=-\delta{\cal L}/\delta {\bf{\cal A}}^3$, we obtain
the eq. (\ref{Lspin}).

\subsection{Electron as a composite object. The vertex operator}
\label{FermiSurface}

The most difficult part of the theory is to find a relation between the true
electron operator
$c_\sigma$ and ``spinned'' fermion $\psi_\sigma$.
%i.e. to find the vertex
%operator ${\cal V}_{\sigma}$ in (\ref{unw}).
The problem is that the local field  theory (\ref{Pauli}),
is written in terms of a not gauge
invariant  operator $\psi_\sigma$. Moreover,
without gauge field (i.e.\ without interaction) $\psi_\sigma$
being a Dirac spinor has 1/2 - orbital
momentum. The electronic states are gauge invariant, and spin singlet
states must have an integer orbital momentum. Another problem is that an
electronic excitation carries a typical momentum of the order of $k_f$,
while typical momenta of  Dirac particles are close to zero.
These difficulties do not
occur while studying the hydrodynamics, but rise in matrix elements.
Below we employ two approaches.  In Appendix we derive the Dirac theory
(\ref{Pauli}) from a microscopical model of a doped Mott insulator and find the
electronic operator and the vertex operator on a regular basis. In this
section we conjecture the form of the electronic operator based on
plausible physical arguments.

The requirements for the electronic states are\\
(i) electron is gauge invariant, i.e. remains unchanged under a
non-singular gradient transformation
${\bf A}\rightarrow {\bf A}+\mbox{\boldmath $\nabla$}\Lambda$;\\
(ii) In a sector with completely filled zero modes, i.e.\ the flux and
the number of
particles obey the topological constraint (\ref{constraint}), a charged
singlet excitation is a
spatial scalar, i.e.\ its wave function has a zero orbital moment $l=0$;\\
(iii) Since an electronic liquid is compressible, the most essential
electronic modes have
momenta $k\sim (2\pi/{\bar\rho})^{1/2}\equiv k_f$.

We find the electronic operator in three steps. First we take care of
the gauge invariance.
 We define
the vertex operator
$V_\sigma({\bf r})$ as an operator which creates a flux quantum
 and a zero mode at the point ${\bf r}$
in the state with the
spin $\sigma=\pm 1/2$\cite{Mand}. It obeys two relations
\begin{eqnarray}
  && V_\sigma^{-1}\mbox{\boldmath $\alpha$}(-i\mbox{\boldmath
$\nabla$}+{\bf A})
   V_\sigma =-i\mbox{\boldmath $\alpha$}\mbox{\boldmath
$\nabla$},
 \label{v1}\\
   &&\,[F({\bf r}'),V_\sigma({\bf r})] = 2\pi V_\sigma({\bf r})\delta({\bf
r}-{\bf r}') .
 \label{v2}
\end{eqnarray}
These  conditions are valid in the subspace of zero modes
(\ref{zm0}) and can be solved by virtue of the commutation relations of
the Chern-Simons theory
$\left[A_x({\bf r}),A_y({\bf r}')\right]= i \pi  \delta({\bf r}-{\bf r}')$.
In terms of the density of particles with spin
$\sigma$: $c_{\sigma}^\dagger c_\sigma=-\mbox{\boldmath $\nabla$}
\cdot {\bf u}_{\sigma}$, or in terms of their
displacements ${\bf u}_{\sigma}$ the vertex operator can be written as
\begin{equation}
 \label{vertex3}
    V_{\sigma}({\bf r})
%\sim e^{ \int \ln\frac{(w-z)}{L}\rho_{\sigma}(z)dzd\bar z}
    =e^{2\pi i\int^{\bf r} {\bf u}_{\sigma}({\bf r}')\times d{\bf r}'
    -2\pi\beta\int^{\bf r}
    {\bf u}_{\sigma}({\bf r}')\cdot d{\bf r}'}  .
\end{equation}
The operator $V^{-1}_\sigma\psi_\sigma$ is gauge invariant.

The second step is the rotational invariance. Let us consider a wave function
of the free Dirac
field in two spatial dimensions.
 In the  basis, where
$\alpha$-matrices are
$\alpha_x=\sigma_3,\,\alpha_y=-\sigma_2,\,\beta=\sigma_1$, the solution
of the Dirac equation with momentum ${\bf p}$  is
$e^{i{\bf p}{\bf r}}e^{-\frac{i}{2}\beta{\rm arg}(\bf
p)}(u_p,\,v_p)$ for the positive energy
$E=+E_p=\sqrt{p^2+\Delta_0^2}$, and
$e^{i{\bf p}{\bf r}}e^{-\frac{i}{2}\beta{\rm arg}(\bf
p)}(v_p,-u_p)$ for $E=-E_p$ ,
where ${\rm arg}(\bf
p)$ is an angle of the momentum ${\bf p}$, relative to the
x-axis and
$u_p=\sqrt{\frac{1}{2}(1+\frac{|{\bf p}|}{E_p})}$ and
$v_p=\sqrt{\frac{1}{2}(1-\frac{|{\bf p}|}{E_p})}$ are the BCS wave
functions. The spinor carries
an angular momentum $l=1/2$.
We  unwind the Dirac field by a chiral rotation
\begin{equation}
 \label{chiral}%
    \psi_\sigma({\bf p})\rightarrow e^{\frac{i}{2}
    \beta {\rm arg}(\bf p)}\psi_\sigma({\bf p}).
\end{equation}%
This singular transformation has a clear
physical sense---it projects the spinor wave function onto a direction of
the momentum ${\bf p}$. Indeed, the chiral transformation (\ref{chiral}) in two
spatial dimensions, where
$\beta=-i/2 \left[\alpha_x,\,\alpha_y\right]$ is equivalent to
 a spatial rotation of the momentum ${\bf p}$ by the angle
${\rm arg}(\bf p)$ which aligns the momentum ${\bf p}$ along the x-axis of
the coordinate system. Now, without topological gauge fluctuations, the
transformed operator
 is a spatial scalar.

The third step is to boost the fermion to the Fermi surface.  In the
chosen basis the upper and the lower components of the Dirac field
$\psi_\sigma=\Big(\psi_\sigma^{(1)},\,\psi_\sigma^{(2)}\Big)$ correspond to
the states propagating forward and backward along the  vector
${\bf p}$. To construct an electronic operator we shift the momentum of the
upper component by the Fermi vector ${\bf k_f}=k_f\frac{{\bf p}}{p}$ directed
along the momentum
${\bf p}$: ${\bf p}\rightarrow {\bf k}\equiv {\bf k_f}+{\bf p}$ and
the momentum of the lower component by
$-{\bf k_f}$: ${\bf p}\rightarrow {\bf k}-2{\bf k_f} = -{\bf k_f}+{\bf p}$
\begin{equation}
 \label{Fermi}
    \left(\begin{array}{c}\tilde c_\sigma({\bf k_f}+{\bf p})
          \\
    \tilde c_\sigma(-{\bf k_f}+{\bf p})\end{array}\right)
   \sim
     e^{\frac{i}{2}\beta {\rm arg}(\bf
k)}\psi_\sigma({\bf p}).
\end{equation}%
 Here we used $\tilde{c}_\sigma$ to indicate that the gauge
field has not been  taken into account yet.

Formally, the chiral rotation (\ref{chiral}) may be also understood in the
following way. The wave function of the Dirac field depends explicitly
on the choice  of $\alpha$-matrices, i.e.\ on the choice of holomorphic
coordinates in a plane. The  chiral transformation aligns  holomorphic
coordinates relative to each point of the ``Fermi surface'', i.e. sets up the
momentum dependent
$\alpha$-matrices
$$\mbox{\boldmath $\alpha$}_{\bf k_f}
 =e^{\frac{i}{2}\beta{\rm arg}({\bf k_f})}\mbox{\boldmath $\alpha$}
e^{-\frac{i}{2}\beta{\rm arg}({\bf k_f})}.$$

In terms of $\tilde c$ the free Dirac Hamiltonian describes isotropic backward
scattering
\begin{equation}
 \label{TP}%
     H=\int \left\{\xi_k\tilde c^\dagger_\sigma ({\bf k}) \tilde c({\bf k})
    +\Delta_0\,\Big( \tilde c^\dagger({\bf k}) \tilde c({\bf k}-2{\bf k_f})
    +h.c.\Big)\right\} \,d{\bf k}  ,
\end{equation}%
where $\xi_k=v(k-k_f)$.

 Assembling all pieces we obtain the relation between the Dirac fermion
and the physical electron in the sector of zero modes
\begin{equation}
 \label{electron}%
    V_\sigma({\bf r}) c_\sigma ({\bf r})\sim \;
      \int e^{-i{\bf k}{\bf r}} e^{-\frac{i}{2} {\rm arg}({\bf
k})}
     \psi_\sigma^{(\beta=-1)} ( {\bf k}-{\bf k_f})\,d{\bf k}
\end{equation}%
where $\psi_\sigma^{(\beta=-1)}$ is the  Dirac field, projected
onto the zero mode sector (
$\frac{1-\beta}{2}
\psi_\sigma= \left(
\begin{array}{c}
     1 \\
    -1
\end{array} \right)
\psi_\sigma^{(\beta=-1)}$).
% and the  vertex operator is
%\begin{equation}
% \label{Ver}%
%   {\cal V}^{-1}_\sigma({\bf r},{\bf k_f})=
%   e^{-\frac{i}{2}\varphi_{{\bf k}}}
%   V^{-1}_\sigma({\bf r})
%\end{equation}%

\section{Matrix Elements  in Topological Liquids}
\label{MEO}

\subsection{Vertex operators}
\label{VO}
To compute matrix elements we will need an operator algebra for the vertex
operator $V_\sigma$
and two additional vertex operators
of the charge and spin sectors
$V=(V_{\uparrow})^{1/2}(V_{\downarrow})^{1/2}$ and
$V_s=(V_{\uparrow})^{1/2}(V_{\downarrow})^{-1/2}$. Below we choose a
holomorphic basis where
$\beta=\sigma_3$ is  diagonal. At $\beta=-1$ (the flux up) the vertex
operators depend on the
holomorphic part of the displacement $u_\sigma=u_{\sigma,x}+u_{\sigma,y}$:
\begin{eqnarray}
   V(w)&\sim &  e^{i\pi \int^w u(z)dz},
 \label{vc} \\
   V_s(w)&\sim &  e^{i\pi\int^w u_{s}(z)dz},
 \label{vs}
\end{eqnarray}
where $u=(u_\uparrow +u_\downarrow)/2;\;\;
u_s=(u_\uparrow -u_\downarrow)/2$.
Due to the commutation relations (\ref{spind}) vertex operators obey the
operator algebra:
\begin{eqnarray}
   V_s(z)c_{\uparrow}(z')   &\sim&
   (\frac{z-z'}{L})^{1/2}c_{\uparrow}(z')V_s(z),
\nonumber\\
   V_s(z)c_{\downarrow}(z')   &\sim&
   (\frac{z-z'}{L})^{-1/2}c_{\downarrow}(z')V_s(z),
\label{OPA} \\
   V(z)c_{\sigma}(z')   &\sim&
   (\frac{z-z'}{L})^{1/2}c_{\sigma}(z')V(z),
\nonumber \\
   V(z)V_s(z')   &\sim & V_s(z')V(z),
\nonumber \\
   V(z)V(z')     &\sim & V(z')V(z).
\nonumber
\end{eqnarray}
Below we also use a two-particle vertex operator
$\mu(u,w)$ which creates a soliton and antisoliton of the spin displacement
\begin{equation}
 \label{vertex1}
   \mu(u,w)=V_s(u)V_s^{-1}(w).
\end{equation}
The operator algebra gives
\begin{eqnarray}
    \mu(u',w') c_{\uparrow}^{\dagger}(u)c_{\downarrow}^{\dagger}(w)
     &=&  \frac{(u'-u)^{1/2}(w'-w)^{1/2}}{(u-w)}
 \nonumber \\
     &\times & c_{\uparrow}^{\dagger}(u)c_{\downarrow}^{\dagger}(w)\mu(u',w')
 \label{OPA1}
\end{eqnarray}
and if the points $u,u'$ and $w,w'$ coincide
\begin{equation}
 \label{OPA2}
   c_{\uparrow}(u)c_{\downarrow}(w)\mu(u,w)
   \sim \frac{a}{(u-w)} \mu(u,w) c_{\uparrow}(u)c_{\downarrow}(w).
\end{equation}

The physical meaning of the vertex operators is straightforward.
The vertex operator of the spin sector $V_s(z)$ creates a soliton of the spin
displacement ${\bf u}_s$ and removes spin up from the site $z$.
Indeed due to (\ref{spind}), we have
\begin{equation}
 \label{soliton}
   \left[V_s(z),{\bf\nabla}{\bf u}_s(z')\right]=V_s(z)\delta(z-z') .
\end{equation}
In contrast, the vertex operator of the charge channel $V(z)$ creates a flux
of the gauge field (the spin chirality), but commutes with displacements
\begin{eqnarray}
      [F(w),V(z)] &=& 2\pi V(z)\delta(w-z),
 \nonumber \\
   \mbox{}  [u(w),V(z)]  &=& 0.
 \label{uVF}
\end{eqnarray}%

Two composite objects $V_s(z)c_\uparrow(z)$ and $V_s(z)c_\downarrow(z)$ are
singlets but carry electric charge. %

\subsection{Matrix elements}
\label{es}

An electronic operator $c_\sigma$ does not 
create an elementary excitation.  The
ground state with an extra particle is created  by a composite  operator
which creates a flux and places a particle into the core of the flux.
Creating a flux also generates spin waves. They are gapful. Let us
first assume that the gap
$\Delta_0$ is very large and  we can neglect spin excitations. 
Then the vertex
operator $V$ of the charge sector  creates a flux and a zero mode which gets
occupied by a particle created by the 
operator ${\psi_\sigma^{(\beta=-1)}}^\dagger$.

Let us start from
the two-particle state with zero momentum:
\begin{eqnarray}
   |N+2\rangle &\sim &  \varepsilon_{\sigma\sigma'}\int
    d{\bf r}\,d{\bf r}'
 \label{ve}  \\
    &\times &  V({\bf r}) V({\bf
r}'){\psi_\sigma^{(\beta=-1)}}^\dagger({\bf r})
   {\psi_{\sigma'}^{(\beta=-1)}}^\dagger({\bf r'})  |N\rangle .
 \nonumber
 \end{eqnarray}

A physical meaning of the composite operator which 
creates the ground state with an extra
particle  is more transparent in terms of gauge invariant
electronic  operators
$c_\sigma\sim V_\sigma^{-1} \psi_\sigma^{(\beta=-1)}$. In these terms
the operators   $V^{-1}\psi_\uparrow^{(\beta=-1)}\sim V_s c_\uparrow$
and
$V^{-1}\psi_\downarrow^{(\beta=-1)}\sim V_s^{-1}c_\downarrow$ consist of the
electron and the vortex (antivortex) of the spin density only. In its turn,
the two-particle state (\ref{ve}) is composed of the two electrons and the
operator (\ref{vertex1}), creating a vortex and
an antivortex of spin density at the points of electron insertions. Electrons
with
opposite spins see each other with vortices of the opposite angular momenta
$l=\pm
1$ (compare to Ref.
\cite{GMFRS90}). This gives an angular momentum $l=2$ for the pair.

With the help of the operator algebra (\ref{OPA},\ref{OPA1},\ref{OPA2}) we
are able
to compute the tunneling amplitude as well as some other matrix elements.

The singlet two-particle matrix element  $\Delta({\bf r}) =\langle N|
c_\uparrow({\bf r})c_\downarrow(0)|N+2\rangle$ has the form
\begin{eqnarray}
     && \Delta({\bf r})\sim
        \int d{\bf k}\, e^{-i\arg ({\bf k})}
       e^{i{\bf k}{\bf r}}
 \nonumber\\
     &&\times \langle N+2|V_s({\bf r})V^{-1}_s(0)
    |N+2\rangle .
 \nonumber
\end{eqnarray}
Thus the two-particle matrix element is given by the correlation function
of the vertex operators
of the spin channel.
The operator algebra 
$V_s(z)V_s^{-1}(z')=\frac{a}{z-z'}:V_s(z)V_s^{-1}(z'):$
implies $\langle N+2|V_s({\bf
r})V^{-1}_s(0)|N+2\rangle\sim \frac{1}{k_fr}e^{-i\arg ({\bf r})}$, so that
\begin{equation}
 \label{two-particle}%
    \Delta({\bf r})\sim\frac{1}{k_fr}e^{-i\arg ({\bf r})}
    \int e^{-i\arg ({\bf k})}   e^{i{\bf k}{\bf r}}d{\bf k}
\end{equation}%

The matrix element
(\ref{two-particle}) is obtained in the limit of a very
large gap $\Delta_0\rightarrow\infty$. This approximation is sufficient in
order to analyze the transformation properties and the 
angular dependence of the
tunneling amplitude. If the gap is not very large, an embedding of 
two extra particles creates a
gapful spinon-antispinon  excitations. These excitations  do 
not interact with the zero mode of the
charged sector and their wave functions (at ${\bf kr}\gg1$) are just the wave
functions of an unperturbed theory (\ref{TP})
\begin{eqnarray}%
     &&\Phi^{(+)}({\bf k},{\bf r})=u_{k-k_f} e^{i{\bf k}{\bf r}}
     +v_{k-k_f} e^{i({\bf k}-2{\bf k_f}){\bf r}},
 \label{positive}\\
     &&\Phi^{(-)}({\bf k},{\bf r})=v_{k-k_f} e^{i{\bf k}{\bf r}}
     -u_{k-k_f} e^{i({\bf k}-2{\bf k_f}){\bf r}}.
 \label{negative}
\end{eqnarray}%
Here the first (second) function corresponds to a positive (negative) energy.
The wave function of a spinon - antispinon pair with
opposite momenta is
\begin{eqnarray}
   \Delta_{\rm BCS}({\bf k},{\bf r})
     = e^{i{\bf kr}} u_{k-k_f}v_{k-k_f} &&
 \label{BCS13} \\
      ={\rm sign}(k-k_f)\, 
      \Phi^{(+)}({\bf k},{\bf r}) {\Phi^{(-)}}^*({\bf k},{\bf r})
     &=& e^{i{\bf k}{\bf r}} \frac{\Delta_0}{\sqrt{\xi_k^2+\Delta_0^2}}.
 \nonumber
\end{eqnarray}%
  Let us notice that in contrast to Cooper's pairing mechanism, the  gap in the
{\it topological mechanism of superconductivity} is
generated via backward scattering and is of the insulator nature.
Nevertheless, the wave function of a singlet spin excitation
with zero relative momentum is the same as the BCS wave function

To obtain matrix elements for a finite gap, we have to replace the factor
$e^{i{\bf k}{\bf r}}$ in eqs.(\ref{two-particle}) by
the spinon-antispinon wave
function (\ref {BCS13})
\begin{equation}
   \label{two-particles-mass}
   \Delta({\bf r})\sim
   e^{-i\arg ({\bf r})}\frac{1}{k_fr}\int  e^{-i\arg ({\bf k})}
   \Delta_{\rm BCS}({\bf k},{\bf r}) d{\bf k}.
\end{equation}

 A state with one extra particle is characterized by two momenta.
One of them is  a momentum
${\bf k}$ of the spin excitation and it is   of the
order of the Fermi momentum. Another one is a small 
momentum  ${\bf q}$ of the zero mode. Arguments similar to the ones used in
the derivation of (\ref{two-particles-mass}) give
\begin{eqnarray}
&&\langle N|c_\uparrow({\bf r})|N+1,{\bf k},{\bf q}\rangle
     \sim  \Phi^{(+)}({\bf k},{\bf r})
 \nonumber \\ 
     &&\times \int    e^{i{\bf q}({\bf r}-{\bf r'})}
    \Big(\frac{e^{-i \arg ({\bf r'})-i\arg ({\bf
k})}}{k_fr'}\Big)^{\frac{1}{2}}d{\bf r'}
\label{tr2mass}
\end{eqnarray}
In momentum representation the one-particle matrix element is
\begin{eqnarray}
   &&\langle N|c_\uparrow({\bf k'})|N+1,{\bf k},{\bf q}\rangle
   \sim \int   d{\bf r'}\,e^{-i{\bf q}{\bf r'}}
   \Big(\frac{e^{-i \arg ({\bf r'})
    -i\arg ({\bf k})}}{k_fr'}\Big)^{\frac{1}{2}}
 \nonumber \\
   && \times  (u_{k-k_f} \delta({\bf k}'-{\bf k}-{\bf q}) +
   v_{k-k_f} \delta({\bf k}'-{\bf k}-{\bf q}-2{\bf k_f}))
 \label{tr3mass}
\end{eqnarray}

These results lead
to a number of important consequences.

\subsection{Orthogonality catastrophe and angle dependence
of the tunneling}
\label{ADT}

(i) {\it The angular dependence of the  pair wave function and of the tunneling
amplitude.}
 The
 pair wave function consists of two vortices of the  same charge---one comes
from the wave function of the zero mode of the
flux  created for a spin up particle by a spin down particle
(the factor $e^{-i\arg ({\bf r})}\frac{a}{r}$ in (\ref{two-particles-mass})).
Another vortex ($
e^{-i\arg ({\bf k})}$) is located at the center of the Fermi surface. The
angular dependence of the pair wave function becomes obvious after the
transformation:
\begin{eqnarray}
   \Delta({\bf r}) &=&  e^{-i2\arg ({\bf r})}\Delta(|{\bf r}|) ,
 \nonumber\\
   \Delta(|{\bf r}|) &\sim&  \frac{1}{k_f r}\int
    e^{i\theta}e^{ikr\cos\theta}
   \frac{\Delta_0}{\sqrt{\xi_k^2+\Delta_0^2}}\,kd k\,d\theta ,
 \label{w}
\end{eqnarray}
where $\theta$ is the angle between ${\bf k}$ and ${\bf r}$.

The  pair wave function forms the d-wave ($l=2$) irreducible complex
representation of  the
rotational group.
Similarly the one particle matrix element carries $l=1$ orbital moment.

(ii) {\it Orthogonality catastrophe.}
The overlap between two ground state wave functions  with $N$ and
$N+2$ electrons does not vanish in a large system.

This is not the case for a one- (or any odd number) particle matrix element. An
attempt to embed a single electron in the system leads to a half-occupied zero
mode state.  As a result this state turns out to be almost orthogonal to all
other states with the same spin and number of particles. The matrix element
(\ref{tr3mass}) vanishes at $q\rightarrow 0$ as
$\langle N|c_\sigma(k+q)|N+1,k,q\rangle\sim q^{1/2}$.
   This is a well-known phenomenon in electronic physics in one dimension
\cite{AndersonOrt}. A {\em new} feature is that it vanishes also as a result of
the phase interference over the Fermi surface  (the integral over
${\rm arg}\,{\bf r}$ in (\ref{tr2mass})).

Due to the general arguments of the previous section, the tunneling
is determined by the instantaneous pair matrix element
(\ref{two-particles-mass}).

(iii){\it Tomographic representation.}
Eq. (\ref{w}) consists of the integral over the Fermi surface and may be
viewed as  a ``tomographic'' representation of the matrix element
\begin{equation}
 \label{w2}
   \Delta({\bf r}) \sim  e^{-i2\arg ({\bf r})}\oint d{\bf k}
   D_{{\bf k}}({\bf r})\int \Delta_{\rm BCS}({\bf k}, {\bf r})kdk
\end{equation}
where the propagator $D_{{\bf k}}({\bf r})$ is given by:
\begin{equation}
 \label{w300}
   D_{{\bf k}}({\bf r})=\frac{k}{{\bf r}\cdot{\bf k}+i{\bf r}\times{\bf
k}}=\frac{1}{r}e^{i(\arg
{\bf k}-\arg{\bf r})}.
\end{equation}
The tomographic representation in  electronic liquids has been anticipated in
\cite{Luther,Anderson}. The new feature is that the relative phase of electron
pairs at different points of the Fermi surface are correlated  by the
factor $\exp({i\arg{\bf k}})$.

(iv){\it Bremsstrahlung.}
It is instructive to rewrite eq. (\ref{w}) in momentum representation
\begin{equation}
 \label{ww}
   \Delta({\bf k}) \sim e^{-i2\arg({\bf k})}\int
   \Delta_{\rm BCS}({\bf k}-{\bf  q})D_{\bf  k}({\bf q})d{\bf q}
\end{equation}
where the propagator
\begin{equation}
 \label{D}
   D_{\bf k}({\bf q})=\frac{k}{{\bf  q}\cdot{\bf k}+i{\bf q}
   \times{\bf k}}=\frac{1}{q}e^{i(\arg {\bf k}-\arg{\bf q})}
\end{equation}
is a holomorphic function relative to ${\bf  k}$, and
\begin{equation}
 \label{D3}
   \Delta_{\rm BCS}({\bf k}) =
   \frac{\Delta_0}{\sqrt{(\epsilon({\bf k})-\mu)^2+\Delta_0^2}}.
\end{equation}
This representation clarifies the physics of the topological superconductor. An
insertion of two particles in the spin singlet state with relative
momentum  ${\bf  k}$ close to ${\bf k}_f$ emits  soft modes of transversal spin
current with the propagator $D_{{\bf k}-{\bf k}_f}({\bf q})$. As a result of
this: (i) ground states differing by an odd number of particles are orthogonal
\cite{Talstra};
(ii) the BCS wave function is dressed by soft transversal spin modes. This
is in line with
1D physics and the {\it bremsstrahlung} effect of QED. The new feature is
the phase of the matrix element of the soft mode (\ref{D}) which is the angle
relative to the Fermi momentum of the electron. By contrast, in the BCS the
density modulations are not individual excitations but are composed from
Cooper pairs. The interaction between density modulations and pairs---the
major effect of the topological mechanism---vanishes in the BCS.

(iv){\it Momentum dependence of the two-particle wave function  and
tunneling amplitude.}
The momentum dependence of the amplitude of the  pair wave function $|\Delta
({\bf k})|$ is drastically different from the BCS (\ref{BCS20}). In the
vicinity of Fermi surface $|k-k_f|\ll k_f$ the integrals (\ref{w2}) can be
computed:
\begin{eqnarray}
  & |\Delta (k)|& -|\Delta (k_f)| \approx
   \frac{\Delta_0}{\epsilon_f}\frac{k_f^2}{k^2}\sinh^{-1}
   \frac{v(k-k_f)}{\Delta_0}    \\
    &\sim &
   \left\{
   \begin{array}{ll}
     v(k-k_f)/\Delta_0 &
     \;\;v|k-k_f| \ll \Delta_0 \\
     {\rm sgn}\,(k-k_f)\log(v|k-k_f|/\Delta_0) & \;\;v|k-k_f| \gg \Delta_0
   \end{array}
   \right.
 \nonumber
\end{eqnarray}
The result gives the universal dependence of the  pair wave function on
$\frac{v(k-k_f)}{\Delta_0}$.
The constant $|\Delta (k_f)|$ is not universal and depends on states far
away from the  Fermi surface.

\begin{figure}
   \centerline{\psfig{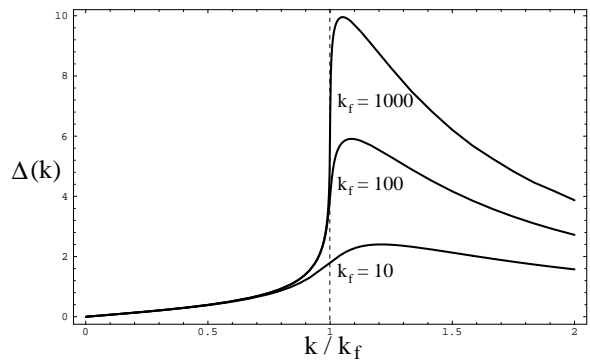}}
   \vspace{0.5 cm}
   \caption{The modulus of  the pair wave function $|\Delta|$ is shown as a
   function of momentum $k$ at $\Delta_0/vk_f=10^{-1},\,10^{-2},\,10^{-3}$.
   In contrast to the BCS gap function,  the pair wave function is asymmetric
   around the Fermi surface ($k/k_f=1$). It is peaked at the scale
   $v(k-k_f)\approx
   \epsilon_f\log^{-1}\frac{\epsilon_f}{\Delta_0}$, which is much bigger
   than $\Delta_0$, and has a long tail away from the Fermi surface.}
 \label{ordpar}
\end{figure}

In contrast to the BCS gap function,  the pair wave function (see
Fig.\ \ref{ordpar}) is
asymmetric around the Fermi surface. It is  peaked at scale
$$v(k-k_f)\approx \epsilon_f
\left(\log\frac{\epsilon_f}{\Delta_0}\right)^{-1},$$
which is much greater than $\Delta_0$. This feature (already emphasized in the
Sec.\ \ref{TOC} where we discussed the frequency dependence) is another
consequence of the  orthogonality catastrophe and is common in one-dimensional
physics \cite{1DHubb}. Also, the pair wave function has a logarithmic
branch cut
at $v(k-k_f)=\pm i\Delta_0$ in contrast to square root singularity of
the BCS function. This indicates that transversal spin current soft modes are
emitted due to the tunneling.

\section{The  pair wave function in the Commensurate Case}
\label{OPCS}

\subsection{The model}

In this section we study the lattice model which
can be derived from the canonical t-J
model of doped Mott insulator under a set of physical
assumptions. This set of assumptions and the derivation is
outlined in Sec.\ \ref{Mott}.

The model describes lattice fermions on a two-dimensional square lattice
interacting with a  gauge field. The number of fermions is
close to the number of lattice sites and the electronic part of the
Hamiltonian is the square of the hopping Hamiltonian
\begin{eqnarray}
 \label{eh11}
   H &=& {\bar t}{\cal H}^2 ,
 \nonumber\\
   {\cal H} &=& \sum_{\langle{\bf a}{\bf b}\rangle}
   \left\{ |{\bf a}\rangle U({\bf a},{\bf b})
   \langle{\bf b}|+|{\bf b}\rangle U({\bf b},{\bf a})
   \langle{\bf a}|\right\}
\end{eqnarray}
with a fluctuating hopping amplitude. Here $|{\bf a}\rangle$ is a site of a
sublattice $A$ of  the square bipartite
lattice, $|{\bf b}\rangle$ is a site of the sublattice $B$, and $\langle{\bf
a}{\bf b}\rangle$ denotes neighboring sites. The phase of a fluctuating
hopping amplitude  is a gauge field $A$
\begin{equation}
 \label{18}
   U({\bf a},{\bf b})=e^{iA_{{\bf a},{\bf b}}}.
\end{equation}
The Hamiltonian (\ref{eh11})  describes the hopping of a particle within
sublattice
$A$ and another particle within sublattice $B$
\begin{eqnarray}
   H &=& {\bar t} \sum_{\langle{\bf a}{\bf a}^{\prime}\rangle,
   \langle{\bf b}{\bf b}^{\prime}\rangle}
   \left\{ \psi^{\dagger}({\bf a})U({\bf a},{\bf a}^{\prime})
   \psi ({\bf a}^{\prime})    \right.
 \label{eh12} \\
   &+& \left. \psi^{\dagger}({\bf b})
   U({\bf b},{\bf b}^{\prime}) \psi ({\bf b}^{\prime})\right\} .
 \nonumber
\end{eqnarray}

The hopping amplitudes  describe two consecutive hops which
leave electrons on the same sublattice---here ${\bf a},{\bf a}^{\prime}$
and ${\bf b},{\bf b}^{\prime}$ are nearest sites of sublattices
$A$ and $B$
\begin{eqnarray}
   U({\bf a},{\bf a}^{\prime})=
   \sum_{\langle{\bf a}{\bf b}\rangle
   \wedge\langle{\bf b}{\bf a}^{\prime}\rangle}
   U({\bf a},{\bf b})U({\bf b},{\bf a}^{\prime}),
 \nonumber\\
   U({\bf b},{\bf b}^{\prime})=
   \sum_{\langle{\bf b}{\bf a}\rangle
   \wedge\langle{\bf a}{\bf b}^{\prime}\rangle}
   U({\bf b},{\bf a})U({\bf a},{\bf b}^{\prime}).
 \label{100}
\end{eqnarray}
The spin of the original electron has been identified with a sublattice, say,
particles with spin up live only on the sublattice $A$ whereas
particles with spin down live on sublattice $B$ (see the Appendix).

It will be explained in the Appendix that the electronic energy achieves
its minimum if the gauge field $A_{{\bf a},{\bf b}}$ forms a flux $\pi$ per
plaquette ({\it flux hypothesis} \cite{chiralphase,WiegmannHasegawa}):
\begin{equation}
\label{200}
{\bar U}_{{\bf a},{\bf a}+{\bf e}_x}
{\bar U}_{{\bf a}+{\bf e}_x,{\bf a}+{\bf e}_x+{\bf e}_y}
{\bar U}_{{\bf a}+{\bf e}_x+{\bf e}_y,{\bf a}+{\bf e}_y}
{\bar U}_{{\bf a}+{\bf e}_y,{\bf a}}=-1 .
\end{equation}

We choose this as a mean field value for the gauge
field $A_{{\bf a},{\bf b}}$ and consider
fluctuations around this mean field. The fluctuations are smooth and slow
and can be
treated in the continuum limit.

In the mean field flux the hopping amplitude along the diagonal of the lattice
cell vanishes ${\bar U}({\bf r},{\bf r}+{\bf e}_x+{\bf e}_y)=0$
due to the maximal interference of the hopping along two different paths
connecting points ${\bf r}$ and ${\bf r}+{\bf e}_x+{\bf e}_y$.
Then the dispersion becomes $\epsilon
({\bf k}) \sim \cos 2k_x+\cos 2k_x$ and the Fermi
energy $\epsilon ({\bf k})=\mu$ at small chemical potential consists of
four small
pockets centered around Dirac points
 ${\bf  k}_{f}\equiv k_{\pm,\pm} = (\pm\frac{\pi}{2},
\pm\frac{\pi}{2})$ (see Fig.\ \ref{tunnampl}).

Let us therefore decompose the fermionic operator
$\psi_\sigma$ into four smooth movers:
\begin{equation}
 \label{cm}
   \psi_\sigma({\bf r}) = \sum_{{\bf k}_{f}}
   e^{i{\bf k}_{f}\cdot{\bf r}} \psi_{{\bf k}_{f},\sigma}({\bf r})
\end{equation}
In the following we refer to the smooth functions $\psi_{
{\bf k}_{f},\sigma}({\bf r})$ as the continuum part  and to the factors
$e^{i{\bf k}_{f}\cdot{\bf r}}$ as the lattice part of the fermion operator
$\psi_\sigma({\bf r})$.

In this basis the mean field Hamiltonian can be written in the continuum
limit as the square of the Dirac operator
\begin{equation}
\label{P1}
H=D^2=(\alpha_x i \partial_x + \alpha_y i \partial_y )^2,
\end{equation}
where $4\times 4$ Dirac matrices $\{\alpha_x,\alpha_y\}=0$ act on four
fermionic modes $\psi_{{\bf
k}_{f}}\equiv \psi_{ \pm\pm}$. The choice of these
matrices (a gauge freedom) corresponds to a relabeling of the Dirac points
and is limited by the symmetry group of the Fermi surface (there are only four
different gauges). We choose Dirac matrices to be
\begin{equation}
 \label{aq}
   \alpha_x=\tau_3\otimes\tau_3,\;\;\;\;
   \alpha_y=\tau_1\otimes\tau_3,
\end{equation}
 where ${\bf \tau}$ are Pauli matrices and the first (second) matrix in
direct product acts on the first  (second) index of the $\psi_{\pm\pm}$. This
choice of matrices corresponds to the lattice Landau gauge and the mean field
lattice hopping amplitudes in this gauge are:
\begin{eqnarray}
   \alpha_x &\equiv & {\bar U} ({\bf r}, {\bf r}+{\bf e}_x)=1,
 \label{lg} \\
   \alpha_y &\equiv & {\bar U} ({\bf r}, {\bf r}+{\bf e}_y)=(-1)^x .
 \nonumber
\end{eqnarray}
The amplitudes of clockwise and counterclockwise diagonal hoppings have
opposite sign:
\begin{eqnarray}
 \label{t22}
   i\beta &=& \alpha_x\alpha_y
   ={\bar U}({\bf r},{\bf r}+{\bf e}_x,{\bf r}+{\bf e}_x+{\bf e}_y)
 \\
   &=& -{\bar U}({\bf r},{\bf r}+{\bf e}_y,{\bf r}+{\bf e}_x+{\bf e}_y)
   =\tau_2\otimes 1=-(-1)^x .
 \nonumber
\end{eqnarray}
The gauge was chosen such that $\alpha_i^\top=-\alpha_i$. The advantage of this
gauge is that electronic modes $c_{\sigma,\pm\pm}$  are indeed slow---the
mean field Hamiltonian (\ref{P1}) achieves the minimal energy at zero
momentum.

Fluctuations of the gauge fields around the mean field value can be easily
incorporated
into the mean field Hamiltonian (\ref{P1}). Setting $\psi({\bf
a})=\psi_\uparrow$ and $\psi({\bf
b})=\psi_\downarrow$, we obtain:
\begin{eqnarray}
 \label{Pauli0un}
   H = \frac{1}{2m}\psi^\dagger_{\sigma}\{
   (i\mbox{\boldmath $\nabla$}-{\bf  A})^2+ \beta F
   \}\psi_{\sigma} .
\end{eqnarray}
Let us note that the diagonal hopping (the last term in (\ref{Pauli0un}))
exists only due to fluctuations. For a more detailed discussion 
see the Appendix.

The Hamiltonian (\ref{Pauli0un})
together with the correspondence between continuum fields and the lattice
(\ref{cm}) is the field theory for the  doped Mott insulator on a bipartite
lattice. To make this theory complete one has to add the relation
between fermionic fields $\psi_{\sigma}$ and original electron $c_{\sigma}$.
This will be done in section Sec.\ \ref{lzms} (see (\ref{ind})).

Doping stabilizes the flux phase by causing the creation
of the flux of the  gauge field in order to absorb extra electrons (see
(\ref{flux1})). This, however, increases the magnetic  energy due to chiral
stiffness (unless  the undoped Mott insulator is a chiral antiferromagnet
\cite{chiralphase}). We assume that at a
certain doping the magnetic energy loses the competition (if any)
with the gain in electronic
energy. Then  a doped flux phase becomes a topological
superconductor
\cite{W1,ZLL,WTS1,WTS2}.

\subsection{Lattice zero modes and lattice vertex operator}
\label{lzms}

To clarify the correspondence  between continuum limit and the lattice, let
us write the ground state  wave function of the mean field Hamiltonian
(\ref{P1}) explicitly. As discussed in the Sec.\ (\ref{FermiSurface}), it is
an eigenfunction of the matrix $\beta$ (\ref{t22}) with eigenvalue $-1$, if
the flux is directed up (i.e., we are adding electrons). This state is doubly
degenerate:
\begin{eqnarray}
   f_{A} & = &  e^{i{\bf  k}_{++}{\bf  r}}
   + e^{i{\bf  k}_{--}{\bf  r}}
   +i(e^{i{\bf  k}_{-+}{\bf  r}}+ e^{i{\bf  k}_{+-}{\bf  r}}),
 \nonumber\\
   f_{B} & = &  e^{i{\bf  k}_{++}{\bf  r}}
   - e^{i{\bf  k}_{--}{\bf  r}}
   +i(e^{i{\bf  k}_{-+}{\bf  r}}- e^{i{\bf  k}_{+-}{\bf  r}}).
 \label{g2bb}
\end{eqnarray}
Solutions are chosen, such that each of them exists on only one sublattice:
$f_A=0$ on sublattice B and $f_B=0$  on sublattice A. The state $f_A$ is
occupied
by an electron with spin up, whereas the state $f_B$ is occupied by an electron
with spin down. These functions are not gauge invariant (i.e.\ depend on the
choice of the  matrices (\ref{aq},\ref{lg},\ref{t22})). To construct a gauge
invariant wave function of the electron one must multiply it by the lattice
Dirac tail:
\begin{equation}
 \label{dt}
   \Psi_\sigma({\bf r},C)={\bar U}({\bf r},C)f_{\sigma}({\bf r}),\;\;\;
   {\bar U}({\bf r},C)=\prod_C\alpha_i ,
\end{equation}
where the product goes over some lattice contour $C$. The wave function
explicitly depends on the contour.

Then the electron operator consists of the vertex operator
$V_\sigma({\bf r})$ (see (\ref{vertex3})), the spinned fermion operator
and the
lattice wave function $\Psi_\sigma$:
\begin{equation}
 \label{ind}
     c_\sigma({\bf r})=
     V^{-1}_\sigma({\bf r})\Psi_\sigma({\bf r},C)\psi_\sigma({\bf r}).
\end{equation}
The factor $\Psi_\sigma({\bf r},C)$ may be considered as the lattice
counterpart of the vertex operator. The form of the electronic operator
(\ref{dt},\ref{ind}) is to be compared with similar equations
(\ref{zm0},\ref{electron})---by decomposing $\Psi_\sigma$ into
Fermi modes $e^{i{\bf  k}_{\pm\pm}{\bf  r}}$ we obtain a discrete analog of the
phase factor $e^{i\varphi/2}$ of (\ref{electron}).

The theory  (\ref{Pauli0un}-\ref{ind}) together with a correspondence between
continuum and lattice fields (\ref{cm}) describes the electronic topological
fluid of the doped Mott insulator on a bipartite lattice.

Below we will use the two-particle singlet zero mode state:
\begin{equation}
 \label{lk}
   \Psi({\bf r},{\bf r}')={\bar U}({\bf r},C_{\bf r})
   {\bar U}({\bf r}',C_{\bf r}')f_A({\bf r})f_B({\bf r}').
\end{equation}
It depends on two strings (contours) ending in points ${\bf r}$ and ${\bf r}'$.
Fluctuations of the strings are physical excitations of the pair (not an
artifact of the approach). The presence of string degrees of freedom
is a general feature of gauge theories and becomes
 important when the coupling with matter is strong.
 In the commensurate case strings fall in four groups
within which
$\Phi$ is the same. These groups correspond to the states with pairing from
different Fermi points ${\bf k}_f$ and ${\bf k}'_f$, i.e.\ to a pair with a
total momentum
${\bf P}={\bf k}_f+{\bf k}'_f=(0,0)\equiv (\pm\pi,\pm\pi),
(\pm\pi,0),(0,\pm\pi)$.
The ground state pair wave function  obviously has zero  momentum
 ${\bf P}=0$. A class of strings which gives zero momentum to a pair is
represented by  two contours following each other from some reference
point up to a point ${\bf r}=(x,y)$ and then a single string along the
$y$-axis to $(x,y')$ and then to the point ${\bf r}'=(x',y')$ along the
$x$-axis. In the chosen gauge  this factor is:
\begin{equation}
 \label{UUp}
   {\bar U}({\bf r},C_{\bf r}){\bar U}({\bf r}',C_{\bf r}')=(-1)^{x(y-y')}.
\end{equation}
Then the wave function is translation invariant
\begin{eqnarray}
 \label{hg}
   \Psi({\bf r},{\bf r}')&=&\sin\frac{\pi}{2}(X+Y)+i\sin\frac{\pi}{2}(X-Y)
 \nonumber\\
   &=&\sum_{{\bf k}_f =(\pm\frac{\pi}{2},\pm\frac{\pi}{2})}
   e^{-i\arg({\bf k}_f)}e^{i\,{\bf k}_f{\bf R}} ,
\end{eqnarray}
where $\arg({\bf k}_f)=\frac{\pi}{4}, \frac{3\pi}{4}, \frac{5\pi}{4},
\frac{7\pi}{4}$
and ${\bf R}={\bf r}-{\bf r}'\equiv (X,Y)$.
This is a direct discrete analog of the factor $e^{i\theta+ik r\cos\theta}$
in Eq. (\ref{w}). The two-particle wave function forms an irreducible
complex
p-wave representation of the crystal group:  under $n\pi/2$ rotation it
produces
the factor $e^{-in\pi/2}$ for $n=1,2,3,4$. It can be viewed as a discrete
vortex located at the center of the Brillouin z\^one.

\subsection{The pair wave function}
\label{copc}

Let us now proceed with matrix elements. The calculations are very similar  to
those of Sec.\ \ref{es}, and we review them briefly.
Let us add two particles in a singlet state with momenta ${\bf k}_f\pm {\bf p}$
symmetric with respect to the center of the small Fermi surface
or equivalently with momenta $\pm({\bf k}_f+{\bf p})$ symmetric with respect to
the center of the Brillouin z\^one. The ground state with $N+2$ particles has
the form:
\begin{eqnarray}
   |N+2\rangle & = & \varepsilon_{\sigma\sigma'}\sum_{{k}_f
   =(\pm\frac{\pi}{2},\pm\frac{\pi}{2})}e^{i{\bf k}_f({\bf r}-{\bf r}')}
   \int d{\bf r}\,d{\bf r}'\,d{\bf p}
 \nonumber  \\
   & & \Psi({\bf r},{\bf r}')\Delta_{\rm BCS}({\bf p},{\bf r}-{\bf r}')
 \label{N}  \\
   &\times & \mu({\bf r},{\bf r}')
   c_{\sigma}^\dagger ({\bf k}_f+{\bf p})
   c_{\sigma}^\dagger(-{\bf k}_f-{\bf p})\;
   |N\rangle  ,
 \nonumber
\end{eqnarray}
where  operator $\mu({\bf r},{\bf r}')$---the creation operator of the
spin vortices with opposite momenta---is given by (\ref{vertex1}), $\Psi({\bf
r},{\bf r}')$ is the wave function of a singlet in the zero mode state
(\ref{hg}), and $\Delta_{\rm BCS}({\bf p},{\bf r})$ is the BCS wave
function (\ref{BCS13}):
The integral over ${\bf }p$ goes over a small Fermi surface (pocket).  Then
the  pair wave function
in terms of Fermi  components (\ref{cm}) has the form:
\begin{eqnarray}
   \Delta({\bf r}_1-{\bf r}_2)
    &\sim& \langle N|c_{\uparrow}({\bf r}_1)
   c_{\downarrow}({\bf r}_2)
   \sum_{{k}_f =(\pm\frac{\pi}{2},\pm\frac{\pi}{2})}
   \int d{\bf r}d{\bf r}'  \nonumber \\
    \Psi({\bf r},{\bf r}') && \Delta_{\rm BCS}({\bf k}_f,{\bf r}-{\bf r}')
       \mu({\bf r},{\bf r}') c_{\sigma}^\dagger ({\bf k})
       c_{\sigma'}^\dagger(-{\bf k}) \;
       |N\rangle
 \nonumber\\
    \sim  \Psi({\bf r}_1,{\bf r}_2) &&
   \frac{e^{i{\rm arg}\,({\bf r}_1-{\bf r}_2)}}{|{\bf r}_1-{\bf r}_2|}
   \sum_{{\bf k}_f=(\pm\frac{\pi}{2},\pm\frac{\pi}{2})}
   e^{i{\bf k}_f({\bf r}_1-{\bf r}_2)}
\nonumber \\
   &\times & \Delta_{\rm BCS}({\bf k}_f,{\bf r}_1-{\bf r_2})
    \langle N|\mu({\bf r}_1,{\bf r}_2)|N\rangle.
 \nonumber
\end{eqnarray}
Using eq.\ (\ref{hg}) we obtain the  pair wave function in two equivalent forms
\begin{eqnarray}
   \Delta({\bf R}) \sim &&
   \frac{\sin\frac{\pi}{2}(X+Y)+i\sin\frac{\pi}{2}(X-Y)}{X+iY}
   \Delta_{\rm BCS}({\bf R})
 \label{kl} \\
   =  \frac{1}{X+iY} && \sum_{{k}_f
   =(\pm\frac{\pi}{2},\pm\frac{\pi}{2})}
   e^{-i\arg({\bf k}_f)}e^{i\,{\bf k}_f{\bf R}}\Delta_{\rm BCS}({\bf R}),
 \label{klr}
\end{eqnarray}
where ${\bf R}={\bf r}-{\bf r}'\equiv (X,Y)$.
The numerator of the first  expression (\ref{kl}) is the discrete analog of the
continuous holomorphic function in the denominator.
Under $\pi/2$ rotation it produces the factor $e^{-i\pi/2}$. Another
$e^{-i\pi/2}$
factor is produced by the continuum part. Both phases add to $e^{-i\pi}=-1$, so
that the tunneling amplitude belongs to the irreducible d-wave
representation  of
the group of rotations of square lattice
\begin{equation}
 \label{dwtr}
   \Delta(-Y,X)=-\Delta(X,Y).
\end{equation}

The second expression (\ref{klr}) is the commensurate version of the
eq.\ (\ref{w}). Let us  look at the tunneling amplitude in momentum space.
It is:
\begin{eqnarray}
 \label{popo}
   \Delta({\bf k}) = \sum_{{k}_f
   =(\pm\frac{\pi}{2},\pm\frac{\pi}{2})}
   e^{-i\arg({\bf k}_f)-i\arg({\bf k}-{\bf k}_f)} f(|{\bf k}-{\bf k}_f|),
\end{eqnarray}
where $\arg({\bf k}_f)=\frac{\pi}{4}, \frac{3\pi}{4}, \frac{5\pi}{4},
\frac{7\pi}{4}$ and $f(p)$ is a smooth function. The tunneling
amplitude  consists of two similarly oriented vortices: one in the center
of the Brillouin z\^one (lattice part) while another is at a Fermi point.
Both of them contribute to the phase (Fig.\ \ref{tunnampl})
\begin{equation}
\label{pl}
   \Delta(-k_y,k_x)=-\Delta(k_x,k_y).
\end{equation}

The  pair wave function in both commensurate and incommensurate
(eqs. (\ref{ww}),(\ref{popo})) case may be written in a unified form
(\ref{w20},\ref{BCS20},\ref{w3})
with a general dispersion $\epsilon_{\bf k}$. In the
incommensurate case of Sec.\ \ref{ADT}
$\epsilon({\bf k})-\mu\sim v_{0}(|{\bf k}|-k_f)$ and the Fermi surface is a
circle. In the commensurate case of the Sec.\ \ref{OPCS} $\epsilon({\bf k}
+(\pm\pi,0))=\epsilon({\bf k}+(0,\pm\pi))$ and has a maximum (minimum) at
${\bf k}=(\pm\pi/2,\pm\pi/2)$. If the chemical potential is close to the
maximum (minimum) of $\epsilon({\bf k})$, the Fermi surface consists of four
pockets with dispersion $\epsilon({\bf k})-\mu\sim \sum_{{k}_f
=(\pm\frac{\pi}{2},\pm\frac{\pi}{2})}v_{0} (|{\bf k}|-k_f)$.

\section{Discussion}
\label{Discussion}

{\bf 1.} {\it Electrons dressed by soft modes of density modulation.}
The models of topological fluids have a lot in common with electronic physics
of one spatial dimension. At the core of this physics are zero modes (or
axial current anomaly) and the orthogonality catastrophe.
Neither electrons nor Cooper pairs are elementary excitations in topological
fluids. The one-electron (any non-singlet state) insertion drastically
changes the ground state of the system, so that the matrix element between
two ground states with N and N+1 electrons vanishes in a macroscopic
system. The matrix elements of two particles in a singlet state do not vanish
but are significantly modified by the interaction with  soft modes of density
modulations. In particular, the poles of Green functions on the mass shell
are replaced by branch cuts. All of this is known in one-dimensional physics,
but now we see similar features in  dimensions greater than one. One of the
reasons for this is that the emission of a soft mode is asymptotically
forward: it does not push an electron to another Fermi point (one can see that
the phase factor in eq.\ (\ref{ww}) suppresses the scattering channels
different from forward scattering).

Eq.\ (\ref{ww})  clarifies the physics of a topological superconductor. A
pair of electrons  emits soft transversal spin modes and interacts with them
(compare with one-dimensional systems \cite{1DHubb1}).  These soft modes also
exist in the BCS theory, but their interaction with electrons vanishes at the
mean field level and is negligible beyond the mean field.

{\bf 2.} {\it Excitations are solitons in charge and spin sectors.}
Similar to one-dimensional electronic liquids the elementary excitations of
topological electronic liquids are solitons of the spin and charge distortions
${\bf u}_s$ and ${\bf u}$. They carry spin 1/2 and no electric charge and
charge 1 and no spin respectively.  The vertex operators  $V_s$ and $V$ of
the Sec.\ (\ref{VO}) are the creation operators of these
particles. An interaction
between solitons of different sectors vanishes at small momenta, so the spin
and charge channels are practically {\it separated}.

{\bf 3.} {\it Complex d-wave  pair wave function with
no nodes in the gap function.}
A new feature of two spatial dimensions  (in addition to the very existence of
superconductivity) is that not only the amplitude but also the phase of the
two-particle matrix element  is strongly modified. The  pair wave function
acquires a phase which is given by $\Delta({\bf r})\sim e^{-2i\arg {\bf
r}}|\Delta({\bf r})|$ and realizes the complex d-wave representation of the
group of spatial rotations. The physics of this d-wave state is drastically
different from the conventional d-wave. The amplitude
of the pair wave function as well as the gap function has no nodes.

{\bf 4.} {\it Asymmetric pair.}
The interaction with soft modes gives rise to a broad
structure of the amplitude of the pair wave function in momentum space around
what one may call Fermi surface. The width of this structure is of the
order of $k_f/\log(\epsilon_f/\Delta_0)$ which is much bigger than
$\Delta_0/v$, the width of the peak of the BCS wave function. It is rather
straightforward to take into account the frequency dependence of the
two-particle matrix element. In order to do it one must
replace
$\Delta_{\rm BCS}$ and
$D_{{\bf k}}({\bf q})$ in the eq. (\ref{ww}) by
\begin{eqnarray}
   \Delta_{\rm BCS}(\omega,{\bf k}) &=&
   \frac{\Delta_0}{\omega^2-(\epsilon({\bf k})-\mu)^2-\Delta_0^2},
 \label{D30} \\
   D_{\bf k}(\omega,{\bf q}) &=& \frac{1}{\omega^2-v^2q^2}
   e^{i(\arg {\bf k}-\arg{\bf q})},
 \label{D12}
\end{eqnarray}
so that
\begin{eqnarray}
   \Delta(\Omega,{\bf k}) &\sim&  e^{-i2\arg({\bf k})}\int d{\bf q}d\omega
 \nonumber \\
   &\times & \Delta_{\rm BCS}(\Omega-\omega,{\bf k}-{\bf  q})
   D_{\bf  k}(\omega,{\bf q}).
 \label{wwOmega}
\end{eqnarray}

The orthogonality catastrophe and a strong interaction with soft modes (in
both spin and charge fluids) also affect  other matrix elements. In
particular, the one-particle correlation function acquires a branch cut on
the mass shell
$\omega=\sqrt{(\epsilon({\bf k})-\mu)^2+\Delta_0^2}$ rather than a pole. It
leads to a broad structure around the Fermi surface in   the photoemission
spectrum,  and to the  distribution of the number of particles, similar to
those known in one dimension
\cite{1DHubb}. It also affects the {\it I-V} characteristics of direct
tunneling (\ref{IV1},\ref{IV2}).

The vertex operator algebra developed in this paper allows one to compute the
long time, large distance behavior of the most interesting matrix elements and
correlation functions.

{\bf 5.} {\it Parity and time reversal symmetry breaking.}
There have been  misconceptions in the literature regarding parity symmetry
breaking of the ground state of two-dimensional topological liquid. The
following comments seem to be in order. The spatial parity and
time reversal symmetry are simultaneously broken in the ground state. This
reflects the chiral nature of zero modes. However it is not
easy to detect this symmetry breaking experimentally. The reason is that,
although the time reversal symmetry is broken, there are no spontaneous
local electric
currents neither in the bulk nor on the edge in any steady state. One can see
it from the hydrodynamics of the charge sector (\ref{Lcharge}), but
this fact remains valid even beyond low frequency range. Even more general,
all diagonal
singlet matrix elements are parity and time reversal even. Therefore, one
should not expect to observe the time reversal symmetry breaking by measuring
Faraday rotation \cite{Faraday} and muon spin relaxation \cite{muon}.

Another matter are off-diagonal or non-singlet matrix elements. The broken time
reversal symmetry is explicit in the complex d-wave tunneling amplitude.
Another manifestations of the broken symmetry can be seen in the spin sector.
Among them is an expectation value of the spin chirality
${\bf S}\cdot{\bf \nabla}{\bf S}\times {\bf \nabla}{\bf S}$
and a novel feature---{\it edge spin current}.
This is a two-dimensional version of the known phenomenon in 1D. A spin chain
with gapful bulk spin excitations develops gapless spin excitations at the
edges. In two-dimensional spin liquid edge excitations are chiral spin
currents. Edge magnetic excitations have been observed in spin chains
\cite{HKAHR90}. One may expect to find these soft edge spin excitations
in model systems with an enhanced boundary (say an array of superconducting
islands).

Another obstacle to observe time reversal symmetry breaking is that
 parity  alternates between odd and even layers \cite{LZL,WTS1}. The genuine
parity breaking takes place only in systems with odd number of layers.
Any realistic junction, however averages over many layers.  The tunneling
amplitude averaged over two layers is
$\Delta \sim \cos(2\arg{\bf k}+\alpha)$ where $2\alpha$ is a relative phase
between the  layers. This phase is determined by the intra-layer crystal
anisotropy and most likely is locked to be zero. In this case the phase of
the pair wave function is  indistinguishable from the conventional d-wave. In a
hypothetical mono-layer tri-crystal type of experiment one would expect the
trapped flux to be an integer (in contrast to a half integer for conventional
d-wave \cite{Tsuei}).

{\bf 6.} In addition to the search for different manifestations of
broken time reversal symmetry one may try to test somewhat
more modest predictions of the theory in cuprates. Among them are:
(i) Magnetic  excitations on the edge.
Although the current averaged over layers vanishes, one may try to
detect softened magnetic modes on the edge;
(ii) Asymmetry of the shape of pair wave function $\Delta(k)$ and its broad
structure (see Fig.\ \ref{ordpar}).
(iii) Absence of nodes in the gap function. To the best of our knowledge the
upper bound for the minimal gap (in (110) direction) set by experiments
\cite{AGL96} is about 3\% of the gap in the crystal axis direction.
This is consistent with the gapless spectrum but also leaves some
room for a small gap.

\acknowledgments

We would like to thank A. Larkin, B. Spivak
and J. Talstra  for numerous discussions and L. Radzihovsky
for collaboration on the initial stages of this project.
Also PBW would like to thank Kenzo Ishikawa and JSPS for
the hospitality in Japan where he worked on this project.
AGA was supported by the Hulda B.\ Rotschild Fellowship,
as well as by MRSEC NSF Grant DMR 9400379.
PBW was supported under NSF Grant DMR 9509533.

\appendix
\section*{Doped Mott Insulator as a Topological Fluid}
\label{Mott}

The phenomenological picture described in Sec.\ \ref{TopSec} may be realized
in certain models of the doped Mott insulator. We start from the canonical
Hubbard
model with an infinite on-site Coulomb repulsion or the t-J model on a square
two-dimensional lattice
\begin{equation}
 \label{tJ}
  	H = \sum_{\langle ij\rangle} t_{ij} c_{i\sigma}^\dagger
   c_{j\sigma} +J_{ij} {\bf S}_{i}\cdot {\bf S}_{j} ,
\end{equation}
where
\begin{equation}
  {\bf S}_{i} = \frac{1}{2} c_{i\sigma}^{\dagger}
  \mbox{\boldmath $\sigma$}_{\sigma\sigma'} c_{i\sigma '}
\label{s}
\end{equation}
is the spin operator of an electron at the lattice
site $i$. The hopping amplitude $t_{ij}$ and the antiferromagnetic exchange
amplitude $J_{ij}>0$ connect the nearest neighbors, the total number of
electrons is close to the number of lattice sites: ${ N}_{e} = { N}_{0} (1 -
\delta)$, while a strong Coulomb interaction does not allow doubly
occupied states:
\begin{equation}
 \label{g2}
   n_{i} =\sum_{\sigma =\uparrow,\downarrow} {c}_{i\sigma}^{+}
   c_{i\sigma}^{\phantom +}= 0\quad\rm or\quad 1  \enspace .
\end{equation}
This model does not have distinct scales, which are necessary to separate
the physics of topological fluids from a variety of other physical
phenomena which occur in a correlated electronic system. We, therefore, are
forced to proceed through the {\it adiabatic approximation}, which is not
justified by any small parameter, but correctly captures the physics of
interest.
In fact the simple model (\ref{tJ}) can be generalized to make the
approximation
parametrically correct. We will not concentrate on this, but rather carefully
emphasize a set of assumptions. In the adiabatic approximation one
considers the
hole motion in a {\it slowly} varying spin configuration.
In this case we may use a semi-classical strategy: first find a static
spin configuration which minimizes the energy of the system with a given
doping, and then take into account quantum fluctuations around the static
mean field.

\subsection{Adiabatic approximation}

\subsubsection{Hopping amplitude, chirality operator}

Adiabatic arguments may be applied to a model where  fast processes are
integrated out. In the t-J model as it is, the dynamics of spins is not
adiabatic. The reason is the short distance antiferromagnetic spin
correlation. Each jump of a hole abruptly changes the spin configuration by
flipping a spin on a sublattice and increases the energy of the system by
$\Delta_J\sim J$. However, two consecutive jumps bring a hole to the same
sublattice, so that a spin exchange between two shifted spins may heal a
wounded antiferromagnetic bond. As a result the spin configuration remains
approximately unchanged only in the second (even) order in  the hopping
process.

To estimate an effective hopping Hamiltonian, in which  single jumps (odd
number of sites) are integrated out (a sort of Schrieffer-Wolff
transformation), we assume that the kinetic energy of the hole is smaller
than the exchange energy of the wounded antiferromagnetic bond (the
antiferromagnetic correlation energy). It allows one to consider the hopping
from, say sublattice $A$ to sublattice $B$ as a virtual process. In other
words the matrix elements of the hopping part of the t-J model (\ref{tJ}) at
low energy states are small---an effective  Hamiltonian occurs as a result of
the second order of perturbation theory in $t$.  A quantum spin liquid with a
spin gap $\Delta_J$, where spin-flip excitations are separated from the
ground state, provides favorable conditions for the adiabatic approximation.
It is required that $t\delta \ll \Delta_J$ for the validity of the adiabatic
approximation.

Let us consider a matrix element of the hopping amplitude between nearest
sites $a$ and $a'$ of the sublattice $A$ in the second order of $t$. Let
$|a\sigma\rangle$  be the low energy states of the antiferromagnet with site
$a$ removed and with $\sigma$ being the spin of the ''removed`` electron. Then
the hopping amplitude is:
\begin{equation}
 \label{Delta0}
   \Delta_{\sigma\sigma'}(a,a') = t^{2}\sum_{aa'}\sum_{|b\rangle}
   \frac{\langle a\sigma|c_{a\alpha}c_{b\alpha}^{\dagger}|b\rangle
   \langle
   b|c_{b\beta}c_{a'\beta}^{\dagger}|a'\sigma'\rangle}{E_{0}-E_{b}},
\end{equation}
where the sum goes over intermediate states $|b\rangle$
with  site $b$  removed and $E_{0}$ ($E_{b}$) is the energy of the
ground (excited) state.

The following three approximations follow from strong short range
antiferromagnetism.

(i) First of all one can replace $|a\sigma\rangle$ by $c_{a\sigma}|\{{\bf
S}\}\rangle$, where $|\{{\bf S}\}\rangle$ is a low energy state of the undoped
antiferromagnet: the spin configuration of the ground state of the
antiferromagnet with and without removed site is not drastically different in
the vicinity of the removed site.

(ii) Next, we may replace $E_{0}-E_{b}$ by  $-\Delta_J$  and take  the
denominator out of the sum in (\ref{Delta0}). Indeed, intermediate states
$|b\rangle$ which contribute to the sum (\ref{Delta0})  are different from
the ground state by a permutation of spins on sites $a$ and $a'$. Their
typical energy is of the order  of $\Delta_J$. Matrix elements of other
states with low energy vanish. This means that the time which a hole spends
on the sublattice
$B$ is very short, so the two consecutive hopping operators in (\ref{Delta0})
act at the same time. As a result,
\begin{eqnarray}
   \Delta_{\sigma\sigma'}(a,a') &\sim& {t^{2}}\sum_{|b\rangle}
   \frac{\langle \{{\bf S}\}|c_{a\sigma}^{\dagger}
   c_{a\alpha}c_{b\alpha}^{\dagger} |b\rangle\langle b
   |c_{b\beta}c_{a'\beta}^{\dagger}c_{a'\sigma'}
   |\{{\bf S}\}\rangle}{E_0-E_b}
 \nonumber\\
   \approx - \frac{t^{2}}{\Delta_J} &\sum_{b} &
   \langle \{{\bf S}\}|c_{a\sigma}^{\dagger}c_{a\alpha}c_{b\alpha}^{\dagger}
   c_{b\beta}c_{a'\beta}^{\dagger}c_{a'\sigma'}|\{{\bf S}\}\rangle ,
 \label{Delta01}
\end{eqnarray}
where the sum in the last equation goes only over sites that belong to a two
step contour connecting points $a$ and $a'$.

The effective hopping amplitude may be re-expressed entirely through spin
operators, namely through the chirality operator \cite{chiralphase}
\begin{equation}
   \Delta_{\sigma\sigma^{\prime}}({\bf a},{\bf a}^{\prime})	  =
   \frac{t^{2}}{\Delta_J}
   \sum_b W_{\sigma\sigma'}({\bf a},{\bf b},{\bf a}') ,
	\label{Delta}
\end{equation}
where $W({\bf a},{\bf b},{\bf a}')$ is the chirality operator  which
drives a hole around a closed path ${\bf a}\rightarrow  {\bf b}\rightarrow
{\bf a}'\rightarrow {\bf a}$
\begin{eqnarray}
   {\bf W}( {\bf a}, {\bf b}, {\bf a}') &\equiv&
   \langle \{{\bf S}\}|c_{a\sigma}^{\dagger}c_{a\alpha}
   c_{b\alpha}^{\dagger} c_{b\beta}c_{a'\beta}^{\dagger}
   c_{a'\sigma'}|\{{\bf S}\}\rangle
\nonumber\\
   &=& (\frac{1}{2}+\mbox{\boldmath $\sigma$}\cdot{\bf S}_a)
   (\frac{1}{2}+\mbox{\boldmath $\sigma$}\cdot{\bf S}_b) (\frac{1}{2}
   +\mbox{\boldmath $\sigma$}\cdot{\bf S}_{a'}) .
 \label{W}
\end{eqnarray}
The effective Hamiltonian, then has the form \cite{KW}
\begin{eqnarray}
 	 H &=& \sum_{\langle{\bf a}{\bf a}^{\prime}\rangle,
   \langle{\bf b}{\bf b}^{\prime}\rangle}
   \{c_\sigma^{\dagger}({\bf a})\Delta_{\sigma\sigma^{\prime}}({\bf a},
   {\bf a}^{\prime}) c_{\sigma^\prime}( {\bf a}^{\prime})
 \nonumber \\
   &+& c_\sigma^{\dagger}({\bf b})\Delta_{\sigma\sigma^{\prime}}
   ({\bf b},{\bf b}^{\prime}) c_{\sigma^\prime}( {\bf b}^{\prime})\}
 \nonumber \\
   &+& \sum_{\langle{\bf a}{\bf b}\rangle}
  \Delta_{J}\{c^{\dagger}_{\sigma}({\bf a})
  \mbox{\boldmath $\sigma$}_{\sigma\sigma^\prime}
  c_{\sigma^\prime}({\bf a})\cdot{\bf S}_b+c^{\dagger}_{\sigma}({\bf b})
  \mbox{\boldmath $\sigma$}_{\sigma\sigma^\prime} c_{\sigma^\prime}
  ({\bf b})\cdot{\bf S}_a\} \nonumber \\
  &+&
  \sum_{\langle{\bf a}{\bf b}\rangle}J {\bf S}_{ a}\cdot {\bf S}_{b} ,
 \label{effham}
\end{eqnarray}
where ${\bf a},{\bf a}'$ and ${\bf b},{\bf b}'$ are the nearest points of
sublattices $A$ and $B$ respectively.

\subsubsection{Gauge fields}

The hopping Hamiltonian (\ref{effham}) is not of practical use---the
hopping amplitudes $W_{\sigma\sigma^{\prime}}( {\bf a}, {\bf b}, {\bf a}')$
and electronic operators are not independent. The practical way is to
introduce a gauge field and a non gauge invariant fermionic operator instead
of the electronic operator. Then  the phase of the hopping amplitude will be a
flux of the gauge field. Let us rotate all spins to the third axis by the
$U(2)$ matrix $g$
\begin{equation}
 \label{ii}
   \mbox{\boldmath $\sigma$}\cdot{\bf S}=g^{-1}\sigma^3 g
\end{equation}
and introduce fermionic operators
\begin{eqnarray}
 \label{i}
   \psi({\bf a}) &=& g_{\uparrow\sigma}c_\sigma({\bf a}),
 \nonumber\\
   \psi({\bf b}) &=& g_{\downarrow\sigma}c_\sigma({\bf a})
\end{eqnarray}
on sublattices $A$ and $B$. Then the hopping Hamiltonian is expressed through
a non gauge invariant but independent fields $\psi$ and
$U({\bf a},{\bf b})$
\begin{equation}
 \label{eh3}
   {\cal H}=\sum_{\langle{\bf a}{\bf b}\rangle}
   \psi^\dagger({\bf a})U({\bf a},{\bf b}) \psi({\bf b})+\mbox{h.c.}
\end{equation}
where the flux of the field $U({\bf a},{\bf b})$ is the chirality
\begin{eqnarray}
    W( {\bf a}, {\bf b}, {\bf a}') &=& \mbox{tr}{\bf W}
      = U({\bf a},{\bf b})U({\bf b},{\bf a}')
      U({\bf a}',{\bf b})
 \nonumber\\
     & =& \frac{1}{8}+\frac{1}{2}{\bf S}_a\cdot{\bf S}_b
     +\frac{1}{2}{\bf S}_a\cdot{\bf S}_{a'}
     + \frac{1}{2}{\bf S}_{a'}\cdot{\bf S}_b
  \label{ws} \\
    &+& i{\bf S}_a \times  {\bf S}_b \cdot {\bf S}_{a'}
 \nonumber
\end{eqnarray}
This is the effective lattice Hamiltonian of the doped Mott insulator.

\subsubsection{Continuum limit: The flux phase stabilized by doping}

The next step in the adiabatic scheme is to separate the fast fields
from the slow ones and
then proceed to the continuum limit. To do this let us first determine the
mean field value of the hopping amplitude (\ref{Delta}), i.e. a static spin
configuration which minimizes the the energy. Among a variety of local minima
we concentrate on a liquid  state, and discard charge (spin) density waves
which break the  crystal symmetry. The most interesting  picture appears if
the electronic part rather than the exchange part of the Hamiltonian dictates
the character of the magnetic state. This happens if the gain in
renormalized  kinetic energy of dopants is larger than the difference between
the energy of the undoped antiferromagnet (Ne\'{e}l state) and the true
ground state. We assume that the true ground state of the doped
antiferromagnet belongs to the so-called flux phase universality class. This
means that the kinetic energy of holes is essentially zero which gives a gain
of the order of $\delta^2 t^2/J$ when compared with the Ne\'{e}l state. The
difference between the magnetic energy of flux and the Ne\'{e}l state
$\Delta_J$ is of the same order but numerically less then $J$. Comparing
these energy scales we obtain that the flux state becomes energetically
favorable starting at some critical doping
$\delta_c \sim (J/t) \sqrt{\Delta_J/J}$. Although  $J\gg t$ the critical
doping may be numerically small. Similar conclusions were drawn in
\cite{ZLL} although on the basis of a different model. This condition seems
compatible with the adiabatic approximation. The latter requires that the
typical renormalized kinetic energy of holes in the flux phase
$\sim\delta t^2/\Delta_J$ be bigger than the typical energy of spin
excitations (spin gap).

The moduli of hopping amplitudes $\Delta(a,a')$ and $\Delta(b,b')$ are
determined by the competition between electronic and magnetic parts of the
Hamiltonian.  Due to the crystal symmetry:
$|\Delta({\bf r},{\bf r}+{\bf e}_x)|=|\Delta({\bf r},{\bf r}+{\bf e}_y)|$,
and in the liquid phase they do not depend on the lattice site. We treat the
mean field value of the modulus as a   phenomenological constant and later
set it to 1. However the phase of the hopping amplitudes may be inhomogeneous
even in a liquid. The character of electronic processes is very sensitive to
the phase. If the kinetic energy of electrons is greater than the magnetic
energy, the phase must be chosen to minimize the electronic energy at a given
values of moduli of hopping amplitudes. In other words the exchange part of
the Hamiltonian directly governs the {\it dot} product of spins on different
sublattices,  while the {\it cross} product of spins is determined by
electronic processes.

The {\it flux hypothesis} \cite{WiegmannHasegawa} suggests that the electronic
energy achieves its minimum if the chiralities along contours
$W({\bf a},{\bf a}+{\bf e}_x,{\bf a}+2{\bf e}_x)$ and $W({\bf a},{\bf a}+{\bf
e}_y,{\bf a}+2{\bf e}_y)$ are positive, while the relative phase between
amplitudes along two different paths connecting sites on  diagonals of a
crystal cell
\begin{eqnarray}
  		W({\bf a},{\bf a}+{\bf e}_x,{\bf a}+{\bf e}_x+{\bf e}_y)
   &=& |W|e^{i\Phi/2},\;\;\;\; \nonumber \\
   W({\bf a},{\bf a}+{\bf e}_y,{\bf a}+{\bf e}_x+{\bf e}_y)
   &=& |W|e^{-i\Phi/2},
 \label{phase}
\end{eqnarray}
is the same in every crystal cell and is equal:
\begin{equation}
   \Phi=\pi(1-\delta) .
 \label{fh}
\end{equation}
The phase $\Phi/2$ (i.e.\ the interference of different paths) is the flux
through a clockwise oriented closed triangular paths within a plaquette.

Let us note that the mean-field flux (\ref{fh}) does not vanish as the
doping $\delta\rightarrow 0$. We, therefore, set it equal to $\Phi=\pi$ and
take the correction into account in the continuum limit. Then the
hopping amplitudes along diagonals
$\Delta({\bf r},{\bf r}+{\bf e}_x+{\bf e}_y)=2|W|\cos\frac{\Phi}{2}$
vanish in
the mean field approximation.
In other words, the two contributions to the diagonal
hopping amplitude along two different paths have opposite phases and cancel
each
other. It means that chirality of two differently oriented triangular
contours have
opposite sign
$\langle W({\bf a},{\bf a}+{\bf e}_x,{\bf a}+{\bf e}_x
+{\bf e}_y)\rangle
=-\langle W({\bf a},{\bf a}+{\bf e}_y,{\bf a}+{\bf e}_x
+{\bf e}_y)\rangle$.
The flux $\pi$ per plaquette implies anticommutativity of translations of a
holon ${\bf r}\rightarrow {\bf r}+{\bf e}_x\rightarrow {\bf r}+{\bf e}_x$
and  ${\bf r}\rightarrow {\bf r}+{\bf e}_y\rightarrow {\bf r}+{\bf e}_x$:
\begin{equation}
 \label{pipi}
   {\bar U}({\bf r},{\bf r}+{\bf e}_x){\bar U}({\bf r},{\bf r}+{\bf e}_y)+
   {\bar U}({\bf r},{\bf r}+{\bf e}_y){\bar U}({\bf r},{\bf r}+{\bf e}_x)
   =0.
\end{equation}

The Fermi-surface of the mean field state consists of four pockets around
Dirac points ${\bf  k}_{f}\equiv k_{\pm,\pm} = (\pm\frac{\pi}{2},
\pm\frac{\pi}{2})$, so we decompose the electron operator  into four smooth
movers
\begin{equation}
 \label{cm1}
   c_\sigma({\bf r}) = \sum_{k_f}c_{\sigma, \pm\pm}({\bf r})
   e^{i{\bf k}_{f}{\bf r}}.
\end{equation}
In the following we refer to the smooth functions $c_{\sigma,
\pm\pm}({\bf r})$ as the continuum part  and to the factors $e^{i{\bf
k}_{f}{\bf r}}$ as the lattice part of the fermion operator $c_\sigma({\bf
r})$.

In this basis the mean field Hamiltonian can be written in the continuum
limit as a square of Dirac operator
\begin{equation}
 \label{P}
   H={\bar t}D^2={\bar t}(\alpha_x i \partial_x
   + \alpha_y i \partial_y )^2 ,
\end{equation}
where $4\times 4$ Dirac matrices $\{\alpha_x,\alpha_y\}=0$ act in the
space labeled by the four Fermi points  $(\pm\pm)$.

\subsubsection{Fluctuations and the field theory}

Now we are ready to take into account smooth fluctuations of the phase of the
hopping amplitudes (fluctuations of moduli are not that important):
$$U({\bf r},{\bf r}+{\bf e}_i)={\bar U}({\bf r},{\bf r}+{\bf e}_i)
e^{iA_i({\bf r})}
$$
and proceed to the continuum limit.
Then the  hopping amplitudes along the crystal  axes become:
\begin{equation}
   U({\bf r},{\bf r}+{\bf e}_i)U({\bf r}+{\bf e}_i,{\bf r}
   +2{\bf e}_i)=e^{iA_i({\bf r})+iA_i({\bf r}+{\bf e}_i)} ,
\end{equation}
whereas diagonal hopping is only due  to chirality
fluctuations
\begin{eqnarray}
   U({\bf r},{\bf r}+{\bf e}_x)U({\bf r}+{\bf e}_x,{\bf r}
   +{\bf e}_x+{\bf e}_y) \nonumber \\
  + U({\bf r},{\bf r}+{\bf e}_y)U({\bf r}
   +{\bf e}_y,{\bf r}+{\bf e}_x+{\bf e}_y)
\sim F.
\end{eqnarray}
The hopping Hamiltonian has the form
\begin{eqnarray}
 \label{Hopping}
   H &=& -{\bar t}\{\sum_{i=x,y}\psi^\dagger({\bf r}+2{\bf e}_i)
   e^{iA_i({\bf r})+iA_i({\bf r}+{\bf e}_i)}\psi({\bf r})
 \nonumber\\
   &+& \psi^\dagger({\bf r}+{\bf e}_x+{\bf e}_y) { F}({\bf r})\psi({\bf r})
 \nonumber \\
   &+& \psi^\dagger({\bf r}+{\bf e}_x-{\bf e}_y)
     F({\bf r})\psi_\sigma({\bf r})
   +\mbox{h.c.}\}.
\end{eqnarray}
Finally all fields are smooth and slow and we may take the continuum limit
\begin{eqnarray}
\label{Pauli20}
   H = \frac{1}{2m}\psi^\dagger\{(i\mbox{\boldmath $\nabla$}-{\bf  A})^2
   + \beta   F \}\psi .
\end{eqnarray}
We find the
hopping Hamiltonian to be Pauli operator. In order to compute matrix
elements of the physical electronic operators, one must complement the
Hamiltonian by the relation between $\psi$ in the continuum and the
$c_\sigma$ on the lattice (\ref{cm1},\ref{i},\ref{ii}).


\begin{thebibliography}{100}

\bibitem{AW}
A.G. Abanov, P.B. Wiegmann, Phys. Rev. Lett. {\bf 78}, 4103-4106 (1997).


\bibitem{R}
D.S. Rokhsar, Phys. Rev. Lett. {\bf 70}, 493 (1993).


\bibitem{L}
R.B. Laughlin, Physica C: Superconductivity {\bf 234}, 280 (1994).


\bibitem{Jos}
D.A. Wollman, D.J. Van Harlingen, W.C. Lee, D.M. Ginsberg and A.J. Leggett,
Phys. Rev. Lett. {\bf 71}, 2134 (1993);
I. Iguchi, Z. Wen, Phys.Rev.B {\bf  49}, 12388 (1994);
D.A. Brawner, H.R. Ott, Phys.Rev.B {\bf  50}, 6530 (1994);
D.A. Wollman, D.J. Van Harlingen, J. Giapintzakis and D.M. Ginsberg,
Phys. Rev. Lett.{\bf 74}, 797 (1995);
A. Mathai, Y. Gim, R.C. Black, A. Amar and F.C. Wellstood, Phys. Rev. Lett.
{\bf 74}, 4523 (1995).

\bibitem{AndersonOrt}
For a recent discussion  see
P.W. Anderson,	Rev. Mod. Phys. {\bf 6}, No. 5a, 1085 (Special Issue, 1994);
P.W. Anderson,	Phys. Rev. Lett. {\bf  64}, 1839 (1990).

\bibitem{AndersonChak}
S. Chakravarty, P.W. Anderson, Phys. Rev. Lett. {\bf  72}, 3759 (1994).

\bibitem{AmbBar}
V. Ambegaokar, A. Baratoff, Phys. Rev. Lett. {\bf  10}, 486 (1963).

\bibitem{GeshLar}
V.L. Geshkenbein, A.I. Larkin,
Pis'ma Zh. Eksp. Teor. Fiz. {\bf  43}, 306 (1986)
[JETP Lett. 43, 395 (1986)].

\bibitem{Froehlich54}
G. Fr\"{o}hlich, Proc. R. Soc. {\bf A223}, 296-305 (1954).

\bibitem{Peierls}
For review, see,
S.A. Brazovskii, N. Kirova, Sov. Sci. Rev. Sect. A {\bf 5}, 99, Harwood
Academic Publ. (1984);
A.J. Heeger, S. Kivelson, J.R. Schrieffer and W.-P. Su Rev. Mod.
Phys. {\bf 60}, 781 (1988).

\bibitem{L1}
R.B. Laughlin, Phys. Rev. Lett., {\bf 60}, 2677 (1988).

\bibitem{MFappr}
A. Fetter, C. Hanna, R. Laughlin, Phys. Rev. B {\bf  39}, 9679 (1989);
Y.-H. Chen, F. Wilczek, E. Witten, B.I. Halperin, Int. J. Mod. Phys.
{\bf B3}, 1001 (1989).

\bibitem{WTS1}
P. Wiegmann, Progr. Theor. Phys.  {\bf  107}, 243 (1992).

\bibitem{WTS2}
P. Wiegmann in {\it   Field Theory, Topology, and condensed matter
systems}, ed. by H. Geyer (Lect. Notes in Phys. {\bf 456}, 177,
Springer 1995).

\bibitem{ZerModeRef}
The term ''zero mode`` or  ''midgap`` state is usually used for the case when
an additional state lies inside the gap and is separated from  both upper and
lower parts of the continuum spectrum. This situation occurs in some
commensurate cases. Here we use the same name for the states which contribute
to the spectral asymmetry between valence and conduction bands  but may not
be separated from the valence band. See e.g.,
A.J. Niemi, G.W. Semenoff Phys. Rep., {\bf 135}, 99 (1986);
R. Jackiw, in: Current algebra and anomalies (World Scientific,
Singapore, 1985), and references therein.

\bibitem{GlobAnom}
The even degeneracy of a zero mode is a necessary condition of the
invariance of the theory under global gauge transformations.
E. Witten,	Nucl. Phys. {\bf  B223}, 422 (1983).

\bibitem{Laughlin88FQHEhTc}
R.B. Laughlin, Science {\bf 242}, 525 (1988).

\bibitem{FQHEelectrodynamics}
D.-H. Lee, S.-C. Zhang, Phys. Rev. Lett. {\bf 66}, 1220 (1991).

\bibitem{Mand}
compare with a soliton operator in Sine-Gordon model,
S. Mandelstam, Phys. Rev. D {\bf  11}, 3026 (1975).

\bibitem{AharonovCasher}
Y. Aharonov, A. Casher, Phys. Rev. A {\bf 19}, 2461 (1979).

\bibitem{GMFRS90}
The fact that fermions with opposite spins see each other with opposite fluxes
in anyon superfluid has been originally suggested in
S.M. Girvin, A.H. MacDonald, M.P.A. Fisher, S.-J. Rey and J.P. Sethna,
Phys. Rev. Lett.{\bf 65}, 1671 (1990).

\bibitem{Luther}
A. Luther, Phys. Rev. B {\bf  19}, 320 (1979);
A. Luther, Phys. Rev. B {\bf 50}, 11446 (1994).

\bibitem{Anderson}
P.W. Anderson, Phys. Rev. Lett. {\bf  64}, 1839 (1990).

\bibitem{Talstra}
J.C. Talstra, S.P. Strong, P.W. Anderson,
Phys. Rev. Lett. {\bf  74}, 5256 (1995).

\bibitem{1DHubb}
Yong Ren, P.W. Anderson, Phys. Rev. B {\bf  48}, 16662 (1993).

\bibitem{chiralphase}
for a references for a chiral magnetic state see,
Ian Affleck, J. Brad Marston, Phys.Rev. B {\bf 37}, 3774 (1988);
P. Wiegmann, Proc. Nobel 	Symposium 73, Sweden,
Physica Scripta  {\bf 27}, 160 (1989);
D. V.  Khveshchenko, P.B. Wiegmann,
Mod. Phys. Lett. B, {\bf 3}, 1383 (1989);
D. V.  Khveshchenko, P.B. Wiegmann, Phys. Lett. B, {\bf 225}, 279 (1989);
X.G. Wen, F. Wilczek, and A. Zee,  Phys. Rev. B {\bf 39}, 11413 (1990).

\bibitem{WiegmannHasegawa}
D. Hasegawa, P. Lederer, T.M. Rice, P. Wiegmann,
Phys. Rev. Lett. {\bf  63}, 907 (1989).

\bibitem{W1}
P.B. Wiegmann, Phys. Rev. Lett. {\bf 65}, 2070 (1990).

\bibitem{ZLL}
Z. Zou, J.L. Levy, R.B. Laughlin, Phys. Rev. B {\bf  45}, 993 (1992).

\bibitem{1DHubb1}
A. Luther, V.J. Emery, Phys. Rev. Lett. {\bf 33}, 589 (1974);
A. Finkelshtein, Pis'ma Zh. Eksp. Teor. Fiz., {\bf 25}, 83 (1977) [JETP
Letters, {\bf 25}, 73 (1977)];
P. Wiegmann, J. Phys.  {\bf  C11}, 1583 (1978).

\bibitem{Faraday}
S. Spielman {\it et.\ al.\ }, Phys. Rev. B {\bf 45}, 3149-3151 (1992).

\bibitem{muon}
R.F. Kiefl {\it et.\ al.\ }, Phys. Rev. Lett. {\bf 63}, 2136-2139 (1989);
R.F. Kiefl {\it et.\ al.\ }, Phys. Rev. Lett. {\bf 64}, 2082-2085 (1990).

\bibitem{HKAHR90}
M. Hagiwara {\it et.al.\ }, Phys. Rev. Lett. {\bf 65}, 3181 (1990).

\bibitem{LZL}
R. Laughlin, Z. Zou and S. Libby, Nucl. Phys. {\bf B348}, 693, (1991).

\bibitem{Tsuei}
C.C. Tsuei et. al., Phys. Rev. Lett. {\bf 73}, 593 (1994).

\bibitem{AGL96}
For review of experimental situation see,
J.F. Annett, N. Goldenfeld, A.J. Leggett,
{\it Physical properties of high temperature superconductors V},
D.M. Ginsberg (Ed.), (World Scientific, Singapore, 1996).

\bibitem{KW}
D.V. Khveshchenko, P.B. Wiegmann, Phys. Rev. Lett. {\bf  73}, 500 (1994).


\end{thebibliography}
\end{document}